\documentclass[longauth]{aa}

\usepackage{natbib, twoopt}
\usepackage{longtable}
\usepackage[breaklinks=true, colorlinks=true, linkcolor=blue, citecolor=blue]{hyperref} 
\bibpunct{(}{)}{;}{a}{}{,}             
\makeatletter
  \newcommandtwoopt{\citeads}[3][][]{\href{http://adsabs.harvard.edu/abs/#3}%
    {\def\hyper@linkstart##1##2{}%
     \let\hyper@linkend\@empty\citealp[#1][#2]{#3}}}
  \newcommandtwoopt{\citepads}[3][][]{\href{http://adsabs.harvard.edu/abs/#3}%
    {\def\hyper@linkstart##1##2{}%
     \let\hyper@linkend\@empty\citep[#1][#2]{#3}}}
  \newcommandtwoopt{\citetads}[3][][]{\href{http://adsabs.harvard.edu/abs/#3}%
    {\def\hyper@linkstart##1##2{}%
     \let\hyper@linkend\@empty\citet[#1][#2]{#3}}}
  \newcommandtwoopt{\citeyearads}[3][][]%
    {\href{http://adsabs.harvard.edu/abs/#3}
    {\def\hyper@linkstart##1##2{}%
     \let\hyper@linkend\@empty\citeyear[#1][#2]{#3}}}
\makeatother

\usepackage{graphicx}
\usepackage[varg]{txfonts}
\usepackage[inline]{enumitem}
\usepackage{siunitx}
\usepackage{xcolor}
\usepackage{placeins}
\usepackage{multirow}

\newcommand{\JWST}{\textit{JWST}}
\newcommand{\HST}{\textit{HST}}
\newcommand{\Gaia}{\textit{Gaia}}
\newcommand{\UVISTA}{UltraVISTA}
\newcommand{\HSC}{HSC}
\newcommand{\CFHT}{CFHT}

\newcommand{\CWeb}{COSMOS-Web}
\newcommand{\COSMOS}{COSMOS}
\newcommand{\COSMOScat}{COSMOS2025}

\newcommand{\CLAUDS}{CLAUDS}
\newcommand{\HSCSSP}{HSC-SSP}

\newcommand{\FIRE}{FIRE}
\newcommand{\NIHAO}{NIHAO}

\newcommand{\SEpp}{\textsc{SExtractor++}}
\newcommand{\PSFEX}{\textsc{PSFEx}}
\newcommand{\LEPHARE}{\textsc{LePHARE}}
\newcommand{\Cigale}{\textsc{Cigale}}

\newcommand{\zphot}{photo-$z$}

\newcommand{\Prob}[1]{{\rm Pr} \left \lbrace #1 \right \rbrace}

\DeclareSIUnit{\dex}{dex}
\DeclareSIUnit{\Msun}{M_\odot}
\DeclareSIUnit{\mas}{mas}
\DeclareSIUnit{\mag}{mag}
\DeclareSIUnit{\arcsec}{arcsec}
\DeclareSIUnit{\parsec}{pc}
\DeclareSIUnit{\jansky}{Jy}
\DeclareSIUnit{\year}{yr}

\setlist[enumerate,1]{label=(\roman*)}

\begin{document} 

   \title{Clumpiness of galaxies revealed in the near-infrared with \CWeb{}}

   \subtitle{Substructures at $1 < z < 4$ and their link to stellar mass and star formation}

   \author{W. Mercier\inst{\ref{Marseille}}\fnmsep\thanks{\email{wilfried.mercier@lam.fr}}
          \and
          B. S. Kalita\inst{\ref{Kavli Pekin}, \ref{Kavli Tokyo}}\fnmsep\thanks{Kavli Astrophysics Fellow and Boya Fellow}
          \and
          M. Shuntov\inst{\ref{DAWN}, \ref{Bohr}}
          \and
          R. C. Arango-Toro\inst{\ref{Marseille}}
          \and
          O. Ilbert\inst{\ref{Marseille}}
          \and
          L. Tresse\inst{\ref{Marseille}}
          \and
          Y. Dubois\inst{\ref{IAP}}
          \and
          C. Laigle\inst{\ref{IAP}}
          \and
          H. Hatamnia\inst{\ref{Riverside}}
          \and
          N. McMahon\inst{\ref{Rochester}}
          \and
          A. L. Faisst\inst{\ref{Caltech}}
          \and
          I. G. Cox\inst{\ref{Rochester}}
          \and
          M. Trebitsch\inst{\ref{LUX}}
          \and
          L. Michel-Dansac\inst{\ref{Marseille}}
          \and
          S.-Y Yu\inst{\ref{Kavli Tokyo}, \ref{Department of Astronomy, Tokyo}}
          \and
          M. Hirschmann\inst{\ref{EPFL}, \ref{INAF}}
          \and
          M. Huertas-Company\inst{\ref{Canaria}, \ref{La Laguna}, \ref{LERMA}, \ref{Paris-Cite}}
          \and
          A. S. Long\inst{\ref{Seattle}}
          \and
          A. M. Koekemoer\inst{\ref{stsci}}
          \and
          G. Aufort\inst{\ref{IAP}}
          \and
          J. S. W. Lewis\inst{\ref{IAP}}
          \and
          G. Gozaliasl\inst{\ref{Computer Finland}, \ref{Department of physics, Helsinki}}
          \and
          R. M. Rich\inst{\ref{UCL}}
          \and
          J. Rhodes\inst{\ref{Jet propulsion}}
          \and
          H. J. McCracken\inst{\ref{IAP}}
          \and
          C. M. Casey\inst{\ref{Santa Barbara}, \ref{Texas University}, \ref{DAWN}}
          \and
          J. S. Kartaltepe\inst{\ref{Rochester}}
          \and
          B. E. Robertson\inst{\ref{Santa Cruz}}
          \and
          M. Franco\inst{\ref{CEA Paris}}
          \and
          D. Liu\inst{\ref{Purple mountain}}
          \and
          H. B. Akins\inst{\ref{Texas University}}
          \and
          N. Allen\inst{\ref{DAWN}, \ref{Bohr}}
          \and
          S. Toft\inst{\ref{DAWN}, \ref{Bohr}}
          }

   \institute{Aix Marseille Univ, CNRS, CNES, LAM, Marseille, France \label{Marseille}
        \and
        Kavli Institute for Astronomy and Astrophysics, Peking University, Beĳing 100871, People{\textquotesingle}s Republic of China \label{Kavli Pekin}
        \and
        Kavli IPMU (WPI), UTIAS, The University of Tokyo, Kashiwa, Chiba 277-8583, Japan\label{Kavli Tokyo}
        \and
        Cosmic Dawn Center (DAWN), Denmark\label{DAWN}
        \and
        Niels Bohr Institute, University of Copenhagen, Jagtvej 128, 2200
Copenhagen, Denmark\label{Bohr}
        \and
        Institut d’Astrophsyique de Paris, UMR 7095, CNRS, UPMC Univ., Paris VI 98 bis boulevard Arago, Paris, France\label{IAP}
        \and
        Department of Physics and Astronomy, University of California, Riverside, 900 University Avenue, Riverside, CA 92521, USA\label{Riverside}
        \and
        Laboratory for Multiwavelength Astrophysics, School of Physics and Astronomy, Rochester Institute of Technology, 84 Lomb Memorial Drive, Rochester, NY 14623, USA\label{Rochester}
        \and
        Caltech/IPAC, MS 314-6, 1200 E. California Blvd. Pasadena, CA 91125, USA\label{Caltech}
        \and
        LUX, Observatoire de Paris, Université PSL, Sorbonne Université, CNRS, 75014 Paris, France\label{LUX}
        \and
        Department of Astronomy, School of Science, The University of Tokyo, 7-3-1 Hongo, Bunkyo, Tokyo 113-0033, Japan\label{Department of Astronomy, Tokyo}
        \and
        Institute of Physics, GalSpec, Ecole Polytechnique Federale de Lausanne, Observatoire de Sauverny, Chemin Pegasi 51, 1290 Versoix, Switzerland\label{EPFL}
        \and
        INAF, Astronomical Observatory of Trieste, Via Tiepolo 11, 34131 Trieste, Italy\label{INAF}
        \and
        Instituto de Astrofísica de Canarias, C/ Vía Láctea s/n, 38205 La Laguna, Tenerife, Spain\label{Canaria}
        \and
        Departamento de Astrofísica, Universidad de La Laguna, 38200 La Laguna, Tenerife, Spain\label{La Laguna}
        \and
        Observatoire de Paris, LERMA, PSL University, 61 avenue de l’Observatoire, F-75014 Paris, France\label{LERMA}
        \and
        Université Paris-Cité, 5 Rue Thomas Mann, 75014 Paris, France\label{Paris-Cite}
        \and
        Department of Astronomy, The University of Washington, Seattle, WA 98195, USA\label{Seattle}
        \and
        Space Telescope Science Institute, 3700 San Martin Drive, Baltimore, MD 21218, USA\label{stsci}
        \and
        Department of Computer Science, Aalto University, P.O. Box 15400, FI-00076 Espoo, Finland\label{Computer Finland}
        \and
        Department of Physics, University of, P.O. Box 64, FI-00014 Helsinki, Finland\label{Department of physics, Helsinki}
        \and
        Department of Physics and Astronomy, UCLA, PAB 430 Portola Plaza, Box 951547, Los Angeles, CA 90095-1547\label{UCL}
        \and
        Jet Propulsion Laboratory, California Institute of Technology, 4800 Oak Grove Drive, Pasadena, CA 91001, USA\label{Jet propulsion}
        \and
        Department of Physics, University of California, Santa Barbara, Santa Barbara, CA 93106, USA\label{Santa Barbara}
        \and
        The University of Texas at Austin, 2515 Speedway Blvd Stop C1400, Austin, TX 78712, USA\label{Texas University}
        \and
        Department of Astronomy and Astrophysics, University of California, Santa Cruz, 1156 High Street, Santa Cruz, CA 95064, USA\label{Santa Cruz}
        \and
        Université Paris-Saclay, Université Paris Cité, CEA, CNRS, AIM, 91191 Gif-sur-Yvette, France\label{CEA Paris}
        \and
        Purple Mountain Observatory, Chinese Academy of Sciences, 10 Yuanhua Road, Nanjing 210023, China\label{Purple mountain}
    }

   \date{Received; accepted}

 
  \abstract
   {Clumps in the rest-frame UV emission of galaxies have been observed for decades. Since the launch of the James Webb Space Telescope (JWST), a large population is detected in the rest-frame near-infrared (NIR), raising questions about their formation mechanism.}
   {We investigate the presence and properties of NIR over-densities (hereafter substructures) in star-forming and quiescent galaxies at $1 < z < 4$ to understand their link to the evolution of their host galaxy.}
   {We identify substructures in \JWST{}/NIRCam F277W and F444W residual images at a rest-frame wavelength of \SI{1}{\micro\meter}.}
   {The fraction of galaxies with substructures with $M_\star > \SI{e9}{\Msun}$ has been steadily decreasing with cosmic time from 40\% at $z = 4$ to 10\% at $z = 1$. Clumps, the main small substructures in the rest-frame NIR, are the most common type and are much fainter (2\% of the flux) than similar UV clumps in the literature. 
   Nearly all galaxies at the high-mass end of the main sequence (MS), starburst, and green valley regions have substructures.
   However, we do not find substructures in low-mass galaxies in the green valley and red sequence.
   Although massive galaxies on the MS and in the green valley have a 40\% probability of hosting multiple clumps, the majority of clumpy galaxies host only a single clump.
    }
   {The fraction of clumpy galaxies in the rest-frame NIR is determined by the stellar mass and SFR of the host galaxies. 
   Its evolution with redshift is due to galaxies moving towards lower SFRs at $z \lesssim 2$ and the build-up of low-mass galaxies in the green valley and red sequence.
   Based on their spatial distribution in edge-on galaxies, we infer that most of substructures are produced in-situ via disk fragmentation. Galaxy mergers may still play an important role at high stellar masses, especially at low SFR.
   }


   \maketitle

\section{Introduction}
\label{sec:Introduction}

The morphology of galaxies is now known to have evolved tremendously with redshift in terms of size \citep[e.g.][]{Williams_2010, vanderwel2014, Martorano2024, Yang2025}, stellar density \citep[e.g.][]{Williams_2010}, and overall 3D shape \citep[e.g.][]{vanderwell2014_3D, Huertas2024}
With the advent of the Hubble Space Telescope (\HST{}), one striking feature in high-resolution images was the presence of kilo parsec-scale, bright, local over-densities observed in the rest-frame ultra-violet (UV), referred to as clumps \citep[e.g.][]{Elmegreen2005, ForsterSchreiber2011, ForsterSchreiber20112, Guo2015, Guo2018, Ribeiro2017}. Importantly, these clumps were found to be ubiquitous, with fractions of clumpy star-forming galaxies in the rest-frame UV reaching typical values of 25\% at $M_\star \sim 10^{9.5}\,\si{\Msun}$ and going up to 40\% at \SI{e11}{\Msun}, with little variation with redshift for $1 < z < 3$ \citep[e.g.][]{Huertas-Company2020}. Besides, they were originally estimated to be particularly massive \citep[$M_\star \gtrsim \SI{e9}{\Msun}$, e.g.][]{ForsterSchreiber20112} without similarly massive counterparts known in the local Universe \citep[$M_\star \lesssim \SI{e7}{\Msun}$ at $z \sim 0$, e.g.][]{Elmegreen1999}.
This naturally raised questions regarding their origin and fate which have been investigated from the point of view of both observations and simulations. For instance, \citet{Genzel2011} found that UV clumps in star-forming galaxies at $z \sim 2$ reside in regions where gravitational instabilities are likely to take place, thus favoring disk fragmentation as the mechanism behind their formation. This mechanism is also favored by \citet{Guo2015} and \citet{Sattari2023}. However, \citet{Guo2015} shows that minor mergers could be a more plausible scenario for galaxies with intermediate stellar-masses at $z < 1.5$. Furthermore, for the brightest galaxies in the rest-frame UV, \citet{Ribeiro2017} rather considered a major merger origin by arguing that a significant proportion of bright galaxies hosts two clumps of similar luminosity, which is too few compared to the number produced by disk fragmentation inside simulations \citep[typically between five and ten, e.g.][]{Perez2013, Bournaud2014}.
Indeed, simulations from \citet{Perez2013, Bournaud2014, Mayer2016, Tamburello2017, Fensch2021, Renaud2024} have shown that it is possible to form massive clumps via fragmentation in galaxies at $1 \lesssim z \lesssim 3$ without invoking galaxy mergers.
According to \citet{Perez2013} and \citet{Bournaud2014}, clumps with masses around \SI{e9}{\Msun} are sufficiently long-lived to migrate towards the galaxy center in less than \SI{1}{\giga\year} and may be a viable channel for bulge growth, funneling gas inwards and migrating gas outwards \citep[e.g.][]{Bournaud2011, Dubois2012, Xu2024, Yu2025}. However, smaller clumps may be disrupted by feedback processes \citep[e.g.][]{Bournaud2014, Mayer2016} or by strong local shear \citep{Fensch2021}, thus leading to simulations where only small clumps can be formed and migration is not possible \citep[e.g. in the \FIRE{} simulation,][]{Oklopcic2017}.
Observationally, this was confirmed by \citet{Faisst2025} for starbursts at $z < 4$ where they found that the fraction of clumpy galaxies is correlated to the star-formation efficiency and is higher in the starburst regime.
High-resolution observations of clumps in lensed galaxies (e.g. see Fig.\,9 of \citealt{claeyssens_star_2023}) also show that more massive clumps tend to be older (\SI{10}{\mega\year} at $10^{5.5}\,\si{\Msun}$ versus \SI{1}{\giga\year} at \SI{e7}{\Msun}) and more gravitationally bound. 
Besides, \citet{Fensch2021} and \citet{Renaud2024} find that the clumpiness of galaxies is primarily determined by their gas content. In contrast, \citet{Perez2013} indicate that fragmentation is significantly more probable to happen in a multi-phasic interstellar medium with temperatures of the order of \SI{e4}{\kelvin}. This is, at the very least, mildly supported by observations \citep[e.g., see Fig.\,9 of][]{Ubler2019}.

It is important to note that the existence, or at least the prevalence, of massive UV clumps remains a topic of debate due to the spatial resolution limitations of \HST{} in the UV that prevents the resolution of clumps on scales smaller than approximately \SI{1}{\kilo\parsec} for galaxies at $z > 1$ \citep[e.g.][]{Dessauges2017, Dessauges2018, Huertas-Company2020}. A similar debate is currently taking place for clumps detected with \JWST{} in the near infra-red \citep[NIR; e.g,][]{claeyssens_star_2023, Claeyssens2024}.
Thanks to the magnifying power of gravitational lensing, it is possible to resolve fainter and smaller clumps in lensed galaxies down to approximately \SI{100}{\parsec} with \HST{} \citep[e.g.][]{Adamo2013, Dessauges2017, Dessauges2018} and \SI{10}{\parsec} with \JWST{} \citep[\JWST{}; e.g.][]{claeyssens_star_2023, Claeyssens2024}. The key takeaway from these analyses is that massive UV and optical clumps are nearly absent from lensed galaxies, with clump stellar masses around \SI{e7}{\Msun} and as low as \SI{e5}{\Msun}.
This has led to the interpretation that the mass and size of massive clumps may be overestimated because of either blending of smaller clumps \citep{Dessauges2017} or contamination of the underlying stellar and gaseous disks within which they are embedded \citep{Huertas-Company2020}. An alternative interpretation has been proposed, which is noteworthy. According to \citet{Faure2021}, who performed parsec-scale simulations of high-redshift gas-rich galaxies and mock \HST{} observations with and without strong lensing, massive clumps may consist of multiple sub-clumps while being simultaneously gravitationally bound structures, thus reconciling discrepancies between lensed and unlensed observations of clumps.

Until recently, the consensus was that galaxies appear clumpy in the rest-frame UV but smooth in the rest-frame NIR. This led to the interpretation that the UV clumpiness is produced by intense bursts of star formation happening in patches, while the smooth morphologies in the NIR correspond to the bulk of stars spread into disk and bulge components.
This view was, however, lately challenged with the advent of \JWST{}. Indeed, recent studies looking at substructures in the rest-frame NIR have found that clumps at these wavelengths are commonly found in galaxies at $1 \lesssim z \lesssim 4$ \citep[e.g.][]{claeyssens_star_2023, Claeyssens2024, Kalita2024a, Kalita2024, Kalita2025}, with the fraction of clumpy galaxies being as high as roughly 40\% \citep{Kalita2024a}, and that they can contribute up to 20\% to the galaxy host's stellar mass \citep{Kalita2025}. Moreover, \citet{Kalita2024a, Kalita2024} have shown that UV clumps in star-forming galaxies at $z \approx 1$ are typically associated with NIR clumps in more than 80\% of cases, whereas the reverse association occurs in only 30\% of instances. This suggests that there might be a non-negligible population of UV-obscured clumps, missed by previous rest-frame UV studies, which revives the longstanding question of the origin of such clumps. Similarly to UV clumps seen in unlensed galaxies, NIR clumps have stellar masses of the order of \SI{e9}{\Msun} and sizes between \SI{100}{\parsec} and \SI{1}{\kilo\parsec} \citep{Kalita2024}. They tend to be found mostly in star-forming galaxies, though \citet{Kalita2024} found that about 15\% of them are associated with quiescent galaxies.

Nevertheless, existing statistical studies of NIR clumps in unlensed galaxies have focused on relatively narrow redshift ranges \citep[$1 \lesssim z \lesssim 2$, e.g.][]{Kalita2024a, Kalita2024} which limits our interpretation of the evolution of clumps in galaxies. Besides, there has been so far a tendency to look at clump properties mostly in massive star-forming galaxies. There is therefore a need to extend these works over a wider redshift range and across a broader population of galaxies, including at low stellar mass and star formation rate (SFR).
In this paper, we propose to detect and study rest-frame NIR clumps, and more generally substructures, at $1 < z < 4$ in a representative population of both star-forming and quiescent galaxies in the \CWeb{} survey \citep{CaseyCWebSurvey}.
First, we present in Sect.\,\ref{sec:Observations} the observations and the sample selection. Then, we describe in Sect.\,\ref{sec:detection} the method to detect substructures and, in particular, the difference between the so-called optimal and intrinsic approaches. Afterwards, we discuss our results in Sect.\,\ref{sec:Results}, including the evolution of the clumpiness of the galaxy population with redshift (Sect.\,\ref{sec:Results/overall}), the characteristics of the detected substructures (Sect.\,\ref{sec:Results/size separation}), and how their presence correlates with the position of the galaxies within the $M_\star - {\rm SFR}$ plane (Sect.\,\ref{sec:Results/MS}). Finally, we interpret our results in Sect.\,\ref{sec:discussion}, first by considering the evolution of galaxies with redshift in terms of stellar mass and SFR (Sect.\,\ref{sec:discussion/GV and RS}), second by discussing the potential impact of young stellar populations to the rest-frame NIR flux of substructures, especially at high specific star formation rate (sSFR; Sect.\,\ref{sec:discussion/young populations}), third by looking at the characteristics of multi-clump systems (Sect.\,\ref{sec:discussion/multi-clumps}), and lastly by trying to constrain the origin of clumps in different galaxy populations (Sect.\,\ref{sec:discussion/formation pathways}).

In what follows, we use a $\Lambda$CDM cosmology with $H_0 = \SI{70}{\kilo\meter\per\second\per\mega\parsec}$, $\Omega_{\rm m} = 0.3$, and $\Omega_\Lambda = 0.7$. Physical parameters are derived assuming a \citet{Chabrier2003} initial mass function (IMF). The images have a pixel scale of \SI{30}{\mas}. Unless otherwise specified, the values and associated uncertainties presented in the tables and figures correspond to the median, 16th, and 84th percentiles, respectively. The latter two are estimated with 1\,000 bootstrap realizations, each time randomly removing 50\% of the sample under consideration.

\section{Observations}
\label{sec:Observations}

\subsection{COSMOS-Web}
\label{sec:Observations/COSMOS-Web}

This paper uses data from the \CWeb{} survey, which was imaged with \JWST{} over three epochs: January 2023, January 2024, and May 2024.
A complete account can be found in \citet{CaseyCWebSurvey}, in the data reduction paper \citep{Franco2025}, and in the \COSMOScat{} catalog paper \citep{Shuntov2025}. The survey consists of \SI{255}{\hour} of imaging in the four \JWST{}/NIRCam F115W, F150W, F277W, and F444W bands over a contiguous \SI{0.54}{\deg\squared} area of the \COSMOS{} field, with a $5\sigma$ limiting magnitude of $27.2-28.2$~AB mag \citep[see][]{CaseyCWebSurvey}. In what follows, we leave out the complementary \JWST{}/MIRI observations because they
\begin{enumerate*}
    \item cover a smaller fraction of the field,
    \item are shallower, and
    \item have a coarser resolution.
\end{enumerate*}
In addition, we use the following bands in the \COSMOS{} field to derive the spectral energy distribution (SED) of the host galaxies and estimate their physical properties and photometric redshift \citep[\zphot{};][]{Shuntov2025}: 
\begin{enumerate*}
    \item $u$ from the Canada France Hawaii Telescope (\CFHT{}) Large Area $U$-band Deep Survey (\CLAUDS{}, \citealt{Sawicki2019}). 
    \item $g$, $r$, $i$, $z$, and $y$ from the Hyper Suprime-Cam (\HSC{}) Subaru Strategic Program (\HSCSSP{}, \citealt{Aihara2022}).
    \item 12 optical medium bands from the reprocessed Subaru Suprime-Cam images \citep{Taniguchi2007, Taniguchi2015}, 
    \item $Y$, $J$, $H$, and $K_{\rm s}$ broad bands from the \UVISTA{} survey \citep{McCracken2012}, as well as one narrow band centered at \SI{1.18}{\micro\meter}, and
    \item \HST{}/ACS F814W \citep{Koekemoer07}.
\end{enumerate*}

A full account of the measurement of the \zphot{} and physical properties of the host galaxies can be found in \citet{Shuntov2025}. We provide a short description below.
We derive the \zphot{} using the template fitting code \LEPHARE{} \citep{Arnouts2002, Ilbert06} with a set of templates from \citet{BruzualCharlot03} combined with 12 star-formation history (SFH) models (exponentially declining and delayed; see \citealt{Ilbert15}). Each template has 43 ages ranging from \SI{0.05}{\giga\year} to \SI{13.5}{\giga\year} and with three attenuation curves \citep{Calzetti00, Arnouts2013, Salim2018} with $E(B-V)$ between 0 and 1.2. Emission lines were included following recipes in \citet{Schaerer09} and \citet{saito20} with the normalization allowed to vary by a factor of two. Absorption by the intergalactic medium was included following analytical corrections provided by \citet{madau95}. 
\LEPHARE{} provides a redshift likelihood distribution, after marginalizing over all galaxy and dust templates, which we used as a posterior distribution (assuming a flat prior) to derive the median redshift value and associated uncertainties. Furthermore, \LEPHARE{} provides physical parameters such as the stellar mass of the galaxies or their SFR, as well as their uncertainties, which are all derived from their posterior distributions. 
According to \citet{Shuntov2025, Shuntov2024}, within our range of magnitudes\footnote{When combining the \JWST{}/NIRCam F277W and F444W bands (see Sect.\,\ref{sec:Observations/COSMOS-Web/selection}), we find the 10th quantile, median, and 90th quantile equal to 21.36, 23.15, and \SI{24.34}{\mag}, respectively.}, the precision of \zphot{} as measured by the normalized absolute deviation \citep{Ilbert2006} is $\sigma_{\rm NMAD} \approx 0.015$ with about 2\% of outliers and there is negligible bias when compared to the spectroscopic compilation of \citet{Khostovan2025}. This results in \zphot{} uncertainties of approximately 0.05 at $1 < z < 2$ that increase with redshift to an average of 0.1 at $2 < z < 3$ and 0.2 at $3 < z < 4$.

In what follows, we use the physical parameters from \LEPHARE{} as a baseline. Furthermore, we use physical parameters derived with the energy-balance SED fitting code \Cigale{} \citep[][using the same photo-$z$ found by \LEPHARE{}]{Boquien19}. A complete description of the modeling performed with \Cigale{} is presented in \citet{Arango2024}. 
We used \citet{BruzualCharlot03} single stellar population models and a \cite{Calzetti00} attenuation law for which
\begin{enumerate*}
    \item we included the \texttt{skirtor} \citep{Stalevski16} module to model the contribution of AGNs to the SED and
    \item we used the \texttt{sfhNlevels} module with a bursty-continuity prior \citep{Ciesla23, Arango23} to model non-parametric SFHs (see Sect.\,4 of \citealt{Arango2024} regarding the robustness of the SFHs).
\end{enumerate*}
The physical parameters appear consistent between \LEPHARE{} and \Cigale{} with stellar masses from \Cigale{} offset by roughly \SI{0.1}{\dex} with respect to \LEPHARE{} and \SI{-0.05}{\dex} for the SFR. We list in Table\,\ref{table:difference LePhare Cigale} the differences between the two codes in the four redshift bins used in the following analysis.
These differences appear consistent with those derived in Appendix\,F of \citet{Shuntov2024} and, as discussed in \citet{Arango2024}, they may be due to using non-parametric SFHs with \Cigale{} and parametric ones with \LEPHARE{} \citep[see also the discussion in Sect.\,6 of][]{Leja2019}, or different attenuation curves.
Therefore, this warrants the use of \Cigale{} as a consistency check throughout the rest of the analysis, in particular in Sect.\,\ref{sec:Results/MS} and in Appendix\,\ref{appendix:complementary figures}.

\begin{table}
\caption{Differences in stellar mass and star formation rate between \Cigale{} and \LEPHARE{} for galaxies in the sample.}
\centering
\resizebox{0.9\linewidth}{!}{
\begin{tabular}{lrrr}

\hline

Redshift range & $\Delta \log_{10} M_\star$ & $\Delta \log_{10} \rm SFR$ \\
& [\si{\dex}] & [\si{\dex}] \\
\hline
\hline\noalign{\vskip 2pt}

$1.0\hphantom{0} < z < 1.3\hphantom{0}$ & 0.12$^{+0.11}_{-0.20}$ & -0.02$^{+0.39}_{-0.30}$ \\[2pt]
$1.3\hphantom{0} < z < 1.75$ & 0.12$^{+0.13}_{-0.24}$ & -0.04$^{+0.34}_{-0.30}$ \\[2pt]
$1.75 < z < 2.5\hphantom{0}$ & 0.12$^{+0.13}_{-0.22}$ & -0.05$^{+0.26}_{-0.29}$ \\[2pt]
$2.5\hphantom{0} < z < 4.0\hphantom{0}$ & 0.16$^{+0.15}_{-0.12}$ & -0.11$^{+0.22}_{-0.28}$ \\[2pt]

\hline\noalign{\vskip 2pt}

\end{tabular}}

\label{table:difference LePhare Cigale}
{\small\raggedright {\bf Notes:} We compute the difference galaxy per galaxy evaluating the parameters from \Cigale{} with respect to \LEPHARE{}. For each redshift bin, we provide the median difference and uncertainties correspond to the 16th and 84th quantiles. Positive values indicate that the parameters from \Cigale{} are higher than those from \LEPHARE{}.\par}
\end{table}

\subsection{Morphological modeling} 
\label{sec:Observations/morphology}

In this paper, we use the morphology of galaxies derived with \SEpp{} \citep{BertinSE++, KummelSE++} using a bulge-disk decomposition \citep{Shuntov2025}.
First, we fit the high-resolution \HST{} and \JWST{} images to constrain the structural parameters of the bulge and the disk, which are the position of their center, scale radius, ellipticity, and position angle. They are forced to be the same for each \HST{} and \JWST{} band.
Then, we fit for the flux of both components in all available bands while keeping the structural parameters fixed. 
For each band, the model was convolved with the appropriate point spread function (PSF). The PSF models were obtained using \PSFEX{} \citep{Bertin2011} on stars detected in a first \SEpp{} run. The stars were selected based on a full width at half maximum (FWHM) - signal-to-noise ratio (S/N) criterion and were validated by comparing the model with the encircled energy of real stars from the \Gaia{} DR3 catalog. Furthermore, for each galaxy a segmentation map was produced during the \COSMOScat{} catalog creation process which we use to isolate galaxies and identify background regions.

\subsection{Sample selection}
\label{sec:Observations/COSMOS-Web/selection}

We select galaxies from the \COSMOScat{} catalog \citep{Shuntov2025}. To study the properties of substructures visible in the rest-frame NIR, we carry out detections in images at a rest-frame wavelength that is the closest to \SI{1}{\micro\meter}. We choose this rest-frame wavelength for a couple of reasons. First, it allows us to probe substructures in a large redshift window from $z = 1$ to $z = 4$, thus giving us access to galaxies before and after the peak of star formation at $z \approx 2$. Second, the NIR window is much less sensitive to dust attenuation and star-formation than in the UV (though see Sect.\,\ref{sec:discussion/young populations} for a discussion relative to this) and should trace much more precisely the stellar mass distribution in galaxies. For instance, using the modified starburst dust attenuation model from \citet{Noll2005} used in \Cigale{} \citep[see][]{Arango2024}, we find that the attenuation at a rest-frame wavelength of \SI{1}{\micro\meter} is 70\% lower than the attenuation in the $V$-band. For our sample (see below), this leads to 92\% of galaxies with an attenuation at this wavelength below \SI{0.28}{\mag} and 56\% of galaxies with a negligible attenuation.
We restrict the detections to \JWST{}/NIRCam F277W ($1 < z < 2$) and F444W ($2 < z < 4$) due to their comparable depth and spatial resolution (see Table\,1 of \citealt{CaseyCWebSurvey}). Using F115W and F150W for rest-frame NIR detections would be advantageous at lower redshift; however this would necessitate matching the depths of the observations and, more critically, the PSFs. Therefore, we choose not to pursue this approach in order to maintain a homogeneous analysis.

Following the stellar mass completeness limit of the \CWeb{} survey (see Fig.\,24 of \citealt{Shuntov2025}), we apply a stellar mass cut of $\log_{10} M_\star / \si{\Msun} > 9$ on the entire sample, prior to performing any detection, resulting in a stellar-mass complete sample of 48\,486 galaxies. This selection excludes active galactic nuclei that are identified in the \COSMOScat{} catalog by SED template matching (for more details, see Sect.\,6 of \citealt{Shuntov2025}).
To guarantee that galaxies are indeed above the completeness limit, we apply this criterion on the lower bound of the stellar mass derived by \LEPHARE{}.
We note that it is common practice to remove edge-on galaxies \citep[e.g.][]{Kalita2024} because it is assumed that substructures are less likely to be detected in highly inclined systems (due to projection effects) and made worse by dust attenuation. However, the latter point should be minimal in the rest-frame NIR. Following \citet{Kalita2024}, applying a disk axis ratio cut of $q > 0.3$, where $q = b/a$ with $a$ the major axis and $b$ the minor axis, would remove 19\% of the sample.
We checked the impact of inclination on our results by splitting the sample into the following bins: edge-on ($q < 0.34$), mildly inclined ($0.34 \leq q \leq 0.87$), and face-on ($q > 0.87$)\footnote{For a razor-thin disk, using $q = \cos i$ with $i$ the inclination, this corresponds to $i > \SI{70}{\degree}$, $\SI{70}{\degree} < i < \SI{30}{\degree}$, and $i < \SI{30}{\degree}$, respectively}. For edge-on and mildly inclined galaxies, we find results that are consistent with the entire sample. For face-on galaxies we observe an excess of substructure detections at $z \gtrsim 2$ that can be as high as a factor of 1.5.
Given that the presence of substructures should be independent of inclination, this may suggest that the detection is underestimated in non-face-on galaxies, but it is not clear why the effect is not stronger for edge-on galaxies compared to mildly inclined ones. This effect is independent of whether galaxies are star-forming or quiescent.
Importantly, while inclination does change the amplitude of detections, it does not affect the observed trends with redshift, stellar mass, and SFR.
Thus, in what follows, we decided to keep edge-on galaxies and to not apply any correction to the detections based on the inclination of the galaxies. 

\begin{figure*}[hbtp]
    \centering
    \includegraphics[width=\linewidth]{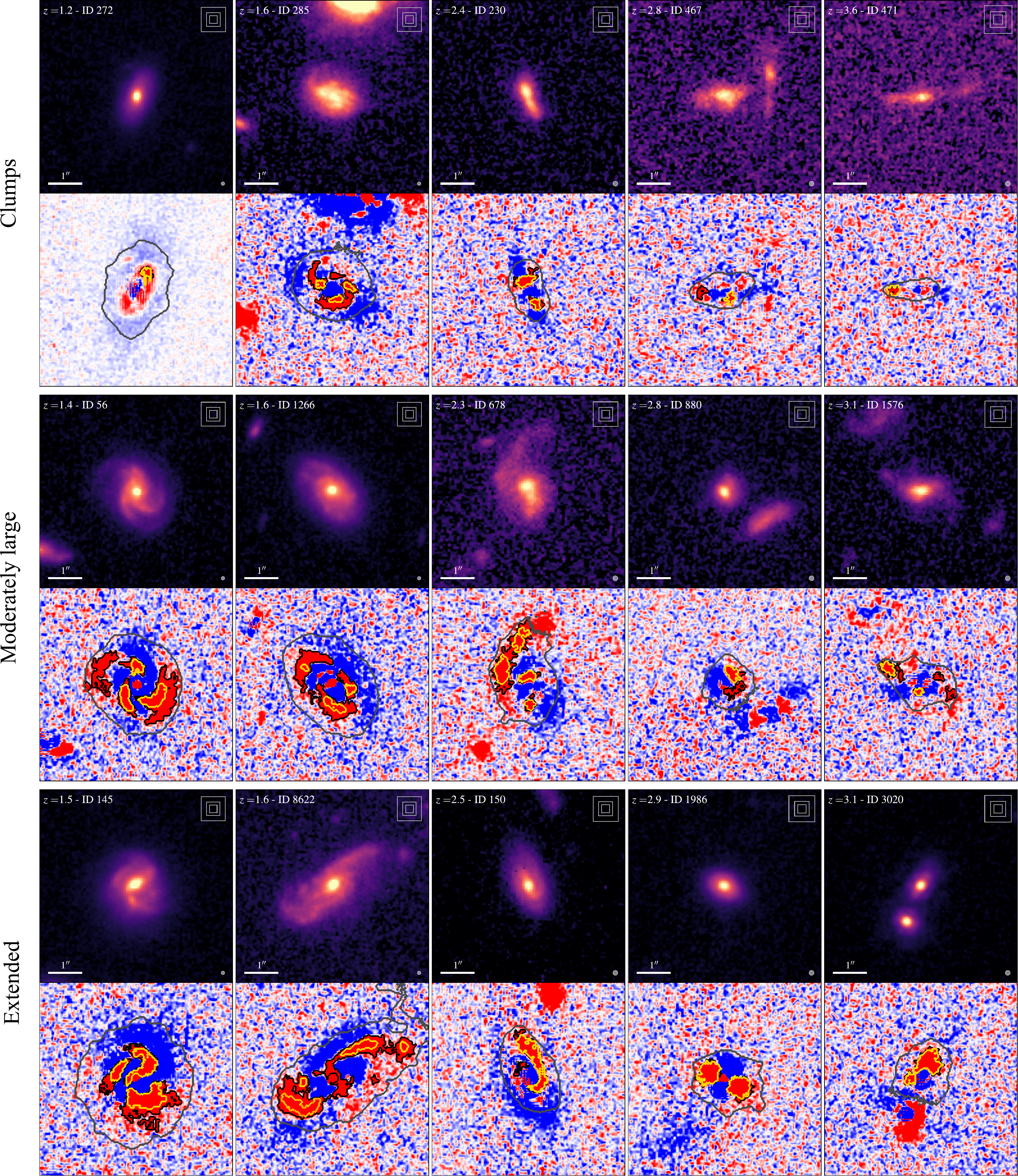}
    \caption{Examples of optimal (black contours) and intrinsic (yellow contours) substructure detections. Galaxies are sorted from left to right by increasing redshift and split into three intrinsic substructure classes: clumps, moderately large, and extended (see Sect.\,\ref{sec:Results/size separation}). For each galaxy, we show the image (F277W for $1 < z < 2$ and F444W for $z > 2$) and its residuals below. Red pixels correspond to positive residuals and blue pixels to negative ones. The scale of the residuals is set to $\pm 2 \sigma$, with $\sigma$ the local background fluctuations. In each image, we show the PSF FWHM at the bottom right and upper limits on the area of substructure classes as gray squares at the top right. We highlight with a gray contour in the residuals the area over which the detection is carried out.}
    \label{fig:example substructures}
\end{figure*}

\section{Detection of substructures}
\label{sec:detection}

\subsection{General overview of the method}
\label{sec:detection/overview}

In this analysis, we define substructures as over-densities with respect to the axisymmetric bulge-disk model fitted onto images, as described in Sect.\,\ref{sec:Observations/morphology}. 
For each galaxy, we detect substructures at a rest-frame wavelength of \SI{1}{\micro\meter} in the residual images in the \JWST{}/NIRCam F277W band for $1 < z < 2$ or in F444W for $2 < z < 4$. We use the segmentation maps produced by \SEpp{} using a combination of all the \JWST{}/NIRCam bands during the \COSMOScat{} catalog creation process to define the extent of the galaxies \citep{Shuntov2025}. We perform the detection in the following two steps.
\begin{enumerate*}
    \item We keep pixels above a flux detection threshold but remove those closer than \SI{1}{\kilo\parsec} from the center and
    \item we combine pixels that form contiguous structures larger than a given area.
\end{enumerate*}
We remove pixels close to the center to avoid detecting drizzling patterns as well as central point-like sources. The choice of \SI{1}{\kilo\parsec} rather than the effective radius of the bulge is motivated by the fact that some galaxies have large bulge sizes. This disfavors the detection of substructures in galaxies with bulge sizes that are overestimated and it also strongly limits our ability to detect substructures in elliptical galaxies. Therefore, a fixed physical size seems more appropriate. From visual inspection, we find that a \SI{1}{\kilo\parsec} exclusion zone is a good compromise to minimize detections near the center while not removing an area that is too extended. As shown in Fig.\,\ref{fig:detection area vs radius}, the removed area is small compared to the detection area of the galaxies which is roughly around 30 and \SI{300}{\kilo\parsec\squared}. 

To set the values of the thresholds, we consider two approaches. The first one, called optimal detection, aims to maximize the detections of substructures whatever the characteristics of the galaxies. However, for a substructure of a specified physical size and intrinsic luminosity, both its apparent size and flux will change with redshift \citep[e.g., see the discussion in][]{Yu2023}. This will lead to a bias where, in particular, intrinsically faint substructures will only be detectable at lower redshifts (see \citealt{Ribeiro2017} for a similar discussion regarding clumps and \citealt{Yu2018} for spiral arms). 
The solution to this problem is to apply thresholds on intrinsic quantities. For the surface area threshold, this implies converting angular scales into physical ones using the angular diameter distance. For the flux threshold, this requires converting to a luminosity using the expression for surface brightness dimming (see Sect.\,\ref{sec:detection/intrisic detection}).

\subsection{Optimal detection approach}
\label{sec:detection/maximal detection}

In this approach, we maximize detections while still striving for high purity. To do so, we compute for each galaxy an optimal flux threshold from the local background using the segmentation maps. The threshold is derived as the standard deviation $\sigma$ of pixels in the background-dominated regions identified by the segmentation map and measured within the residual images. We mask all pixels with a flux below $2\sigma$, which we find to be the ideal threshold to eliminate nearly 99\% of false detections (see Appendix\,\ref{appendix: detection} for a more detailed explanation about this choice).
To limit spurious detections due to background fluctuations that are correlated on PSF scales, we apply a surface area threshold which corresponds to the area enclosed by the FWHM of the PSF in the F444W band (\SI{0.15}{\arcsecond}). Thus, substructures must extend over a contiguous area larger than $\pi \times ({\SI{0.15}{\arcsecond}}/2)^2 = \SI{0.0177}{\arcsec\squared}$ which corresponds to 20 pixels or \SI{1.3}{\kilo\parsec\squared} at $z = 1.5$. In practice, however, substructures with a size close to the PSF FWHM are unresolved and, therefore, we do not try to estimate their intrinsic size. In what follows, we quote the lower bound on the substructure surface to remind the reader that we exclude any substructure less extended than this threshold.

\subsection{Intrinsic detection approach}
\label{sec:detection/intrisic detection}

\begin{table}
\caption{Number of galaxies per bin of substructure area and redshift.}
\centering
\begin{tabular}{lrrr}

\hline
Substructures & \multicolumn{3}{c}{Number of galaxies} \\

\multicolumn{1}{c}{[\si{\kilo\parsec\squared}]} & \multicolumn{1}{c}{$1 < z < 2$} & \multicolumn{1}{c}{$2 < z < 3$} & \multicolumn{1}{c}{$3 < z < 4$} \\
\hline
\hline\noalign{\vskip 2pt}
\multicolumn{4}{c}{\textbf{Optimal detection}} \\[2pt]

$1.13 < S < 4$             & 15\,622 & 6\,552 & 2\,257 \\
$\hphantom{1.}4 < S < 13$ & 11\,092 & 3\,853 & 1\,017 \\
$\hphantom{.}13 < S < 40$ & 5\,730 & 1\,227    & 224 \\[2pt]
$1.3 < S < 40$            & 20\,605 & 8\,610 & 2\,850 \\
\hline\noalign{\vskip 2pt}
\multicolumn{4}{c}{\textbf{Intrinsic detection}} \\[2pt]

$1.3 < S < 4$             & 4\,252 & 3\,500 & 1\,990 \\
$\hphantom{1.}4 < S < 13$ & 2\,080 & 1\,814    & 835 \\
$\hphantom{.}13 < S < 40$ & 434   & 444    & 177 \\[2pt]
$1.3 < S < 40$            & 5\,298 & 4\,600 & 2\,502 \\

\hline\noalign{\vskip 2pt}
 & 26\,000 & 14\,823 & 6\,726 \\

\hline\noalign{\vskip 2pt}

\end{tabular}

\label{table:number of galaxies}
{\small\raggedright {\bf Notes:} The number of galaxies with substructures varies depending on the detection method used. We provide in the last row the total number of galaxies independently of whether a substructure is detected or not. \par}
\end{table}

Here, we take into account the redshift dependency of the surface and flux of substructures. Doing so, we ensure that substructures are detected at a similar intrinsic level independently of their redshift. The flux per unit frequency $S_\nu (\nu_0)$ observed at a frequency $\nu_0$ scales with the luminosity per unit frequency (hereafter specific luminosity) $L_\nu (\nu_1)$ emitted at a frequency $\nu_1$ and with redshift $z$ as \citep{LongairBook}

\begin{equation}
    S_{\nu} (\nu_0) = L_{\nu} (\nu_1) \times (1+z) / \left (4\pi D_{\rm L}^2 \right ),
    \label{eq:flux density}
\end{equation}
where $D_{\rm L}$ is the luminosity distance. The distribution of $L_\nu (\SI{1}{\micro\meter})$ for the $2\sigma$ flux detection levels measured in Sect.\,\ref{sec:detection/maximal detection} is shown in Fig.\,\ref{fig:intrinsic detection threshold}\footnote{We convert $L_\nu$ into AB magnitude times \si{\mega\parsec\squared} with the formula $-2.5 \log_{10} L_\nu + 8.9$ where $L_\nu$ is computed with $S_\nu$ in \si{\jansky} and $D_{\rm{L}}$ in \si{\mega\parsec}.}.
We find that 95\% of the galaxies have a $2\sigma$ luminosity threshold deeper than \SI{5.58}{\mag\mega\parsec\squared} and convert this intrinsic value back into an observed flux using Eq.\,\ref{eq:flux density}:

\begin{equation}
    m_{\rm det} = 8.33 - 2.5 \log_{10} \left [ \frac{1 + z}{D_{\rm L}^2} \right ],
    \label{eq:flux detection curve}
\end{equation}
with $m_{\rm det}$ in AB magnitude and $D_{\rm L}$ the luminosity distance in \si{\mega\parsec}. Here, $m_{\rm det}$ represents the faintest magnitude a substructure must have to be detected at any of the redshifts considered in this analysis\footnote{Add \SI{3.25}{\mag} or divide the corresponding flux by 20 to get the threshold applied at the pixel level.}.

The angular diameter distance reaches its peak at $z \approx 1.6$.
For a surface detection threshold of \SI{0.0177}{\arcsec\squared}, this corresponds to a maximum intrinsic surface of \SI{1.27}{\kilo\parsec\squared} at $z = 1.6$ (versus \SI{1.13}{\kilo\parsec\squared} at $z = 1$ and \SI{0.85}{\kilo\parsec\squared} at $z = 4$). We therefore take this value as the detection limit and convert it back into an observed criterion:

\begin{equation}
    \mathcal{S}_{\rm det} = 0.054 / D_{\rm A}^2,
    \label{eq:surface detection curve}
\end{equation}
with the constant chosen so that $\mathcal{S}_{\rm det}$ is in \si{\arcsec\squared} and the angular diameter distance $D_{\rm A}$ in \si{\giga\parsec}. 
We show in Fig.\,\ref{fig:detection curves}, the two detection curves used for the intrinsic detection, as given by Eqs.\,\ref{eq:flux detection curve} and \ref{eq:surface detection curve}. The net effects are that we allow the detection of:
\begin{enumerate*}
    \item fainter substructures at higher redshift and
    \item smaller substructures at $z \approx 1.6$.
\end{enumerate*}
The numbers of galaxies detected in bins of redshift and substructure area (denoted $S$) for both detections are provided in Table\,\ref{table:number of galaxies}. 
Effectively, the net effect of the intrinsic approach is to reduce the detections of substructures at $z < 4$. Thus, in essence, $z = 4$ will act as the reference redshift for the interpretation of the results presented in Sect.\,\ref{sec:Results}.

\subsection{Sample cleaning and probability of detections}
\label{sec:detection/sample cleaning}

We decided to manually clean the sample and the detections in the intrinsic approach. This seemed necessary since visual inspection showed that some substructures were not real but were obvious artifacts and that there was no easy automatic way to flag them. Such artifacts include
\begin{enumerate*}
    \item drizzling patterns near the bulge for galaxies with bright centers,
    \item positive-negative residuals at the periphery of the bulge induced by a slightly offset center,
    \item wrong inclinations for highly-inclined systems easily recognizable by the residuals they produce along the major and minor axes and
    \item bulge-disk models that failed during fitting.
\end{enumerate*}
We manually removed artifacts (i) and (ii) from the detections and kept the galaxies in the analysis. For artifacts (iii) and (iv), we removed the galaxies from the sample. We also dismissed galaxies whose segmentation map falls entirely over another galaxy since we cannot efficiently identify substructures in such cases. Thus, we cleaned about 2\% of galaxies (artifacts i and ii) and similarly removed 2\% of galaxies from the final sample (artifacts iii and iv).
Examples of substructures detected with the optimal and intrinsic approaches are shown in Fig.\,\ref{fig:example substructures}. For each galaxy, we show the original image (top) and the residuals (bottom) with the contours of the detected substructures overlaid on top. We split examples into three categories: galaxies holding at least one clump ($1.3 < S / \si{\kilo\parsec\squared} < 4$, with $S$ the surface of the substructure), one moderately large substructure ($4 \leq S / \si{\kilo\parsec\squared} < 13$), and one extended substructure ($13 \leq S / \si{\kilo\parsec\squared} \leq 40$) detected by the intrinsic approach (see Sect.\,\ref{sec:Results/size separation} for more details). We note that a galaxy may contain more than one substructure and that a small substructure detected with the intrinsic approach may actually be part of a larger substructure detected with the optimal one.

The chosen size for clumps corresponds to the typical range of sizes measured in the literature \citep[e.g.][]{Guo2015, Guo2018, Kalita2024, Kalita2025}. On the other hand, visual inspection shows that extended substructures mostly correspond to spiral arms and tidal tails (see examples in Fig.\,\ref{fig:example substructures}) and that moderately large substructures act an intermediate class. 
This is confirmed by estimating the shape of substructures for these three categories with two shape metrics. First, we use the circularity parameter defined as $4\pi \mathcal{A} / \mathcal{P}^2$, with $\mathcal{A}$ and $\mathcal{P}$ the area and perimeter of a substructure. Second, we use the ratio of the eigenvalues of the inertia tensor of the flux map of each substructure. Value close to unity mean circular shapes. The median for each category, as well as the 16th and 84h percentiles, are provided in  Table\,\ref{table:substructure shape}. We see that the more extended the substructure, the more elongated it is.
In what follows, we do not consider substructures more extended than \SI{40}{\kilo\parsec\squared} because, as illustrated in Fig.\,\ref{fig:histogram area of substructures}, there are very few of them at $z > 3$. On the other hand, below this value, the distributions of the area of substructures is comparable across the entire redshift range.

Finally, we found during visual inspection cases where substructures fall in-between two or more galaxies, making it challenging to assign them with high confidence to one galaxy or the other. To solve this issue, we have devised a weighting scheme that assigns to each substructure a probability to belong to any galaxy in its neighborhood. This probability can then be used as a weight during the analysis. A full account of this scheme is given in Appendix\,\ref{appendix:weighting scheme}. Its main characteristics are as follows.
First, if the angular separation between a substructure and a galaxy $G_1$ is twice larger than with a second galaxy $G_2$, the substructure has half the probability of belonging to $G_1$ than $G_2$. And second, a substructure must belong to one of the galaxies in its vicinity.
In what follows, we present results with this weighting scheme. We note that we checked the effect of not weighting the detections and find minimal impact. For instance, the fraction of galaxies containing at least one substructure globally increases by 8\% with respect to the values presented in the rest of this paper.

\begin{table}
\centering
    \caption{Measurements of the shape of substructures.}
    \label{table:substructure shape}
    \begin{tabular}{lccc}
    \hline
     & Clump & Moderately large & Extended \\
    \hline
    \hline\noalign{\vskip 2pt}
    Circularity & $0.52^{+0.20}_{-0.20}$ & $0.40^{+0.19}_{-0.17}$ & $0.28^{+0.19}_{-0.12}$ \\[2pt]
    Inertia tensor & $2.6^{+2.87}_{-1.05}$ & $3.52^{+4.25}_{-1.70}$ & $3.90^{+5.15}_{-1.85}$\\[2pt]
    \hline\noalign{\vskip 2pt}
    \end{tabular}
    
    {\small\raggedright {\bf Notes:} Values correspond to the median, 16th, and 84th percentiles. We use two shape measurements. (i) Circularity which is $4\pi \mathcal{A} / \mathcal{P}^2$, where $\mathcal{A}$ is the area of a substructure and $\mathcal{P}$ is its perimeter. And (ii) the ratio of the eigenvalues of the inertia tensor of the image of the substructures. For each metric, values closer to unity correspond to rounder shapes.\par}
\end{table}

\subsection{Completeness of the detections}
\label{sec:detection/completeness}

We consider a substructure encompassing $n$ pixels. Since the luminosity of each of its pixels must be larger than a given threshold (denoted $l_{\rm det}$), the lowest detectable luminosity for a substructure of this size is $n \times l_{\rm det}$. If some pixels are brighter than $l_{\rm det}$, then the substructure will be detected with a higher luminosity. If $k < n$ pixels are fainter than $l_{\rm det}$, then either the substructure will be detected with an area encompassing $n - k$ pixels or, if its area is below the surface detection threshold, the substructure will not be detected. In other terms, using Eq.\,\ref{eq:flux detection curve} we can derive a completeness limit that depends on the area of the substructures. We show in Fig.\,\ref{fig:completeness} the completeness in terms of magnitude at $z = 1$, 2, 3, and 4 as a function of the area of substructures. As expected, larger substructures have a brighter limit. However, we note that the limit in terms of luminosity ($L_\nu$) is independent of redshift. Furthermore, because of the redshift dependence of Eq.\,\ref{eq:flux detection curve}, more distant galaxies have a fainter limit.

We can derive an estimate of the stellar mass completeness limit of the intrinsic detection assuming the $M_\star - L_\nu$ relation at a rest-frame wavelength of \SI{1}{\micro\meter} for substructures follows that of galaxies. We linearly fit the latter using the entire sample\footnote{The best-fit relation found is $\log_{10} M_\star = 1.1 \log_{10} L_\nu + 9.7$ with $M_\star$ in \si{\Msun} and $L_\nu$ in \si{\kilo\jansky\mega\parsec\squared}} and derive the completeness limit shown as the plain dark red line in Fig.\,\ref{fig:completeness}\footnote{It is evaluated at $z = 1.5$. As Fig.\,\ref{fig:completeness} is shown in terms of physical area, there is a small redshift dependence with the zero-point around \SI{0.2}{\dex} higher at $z = 4$.}.
Alternatively, we use the $M_\star - S_\nu$ relation from \citet{Kalita2025} for clumps in star-forming galaxies at $z \approx 1.5$\footnote{This relation is $\log_{10} M_\star = 1.22 \log_{10} S_\nu + 13.35$ with $M_\star$ in \si{\Msun} and $S_\nu$ in \si{\milli\jansky}.}. This is shown as the salmon-colored line in Fig.\,\ref{fig:completeness}.
Both methods agree with a difference of at most \SI{0.5}{\dex} for the most extended substructures. The completeness limit varies from about \SI{e8}{\Msun} for the smallest substructures to nearly \SI{e10}{\Msun} for the largest ones. In particular, this shows we miss "low surface brightness substructures". As an indication, when correcting for surface brightness dimming, this corresponds to a surface brightness limit of about \SI{19.5}{\mag\per\arcsec\squared} at $z = 1$ and \SI{18.1}{\mag\per\arcsec\squared} at $z = 4$ which are both above the historical value of \SI{21.65}{\mag\per\arcsec\squared} used to categorize low surface-brightness galaxies (see \citealt{Boissier2019} for a discussion).

\section{Results}
\label{sec:Results}

\subsection{Global characterization of substructures}
\label{sec:Results/overall}

\begin{figure*}
    \centering
    \includegraphics[width=\linewidth]{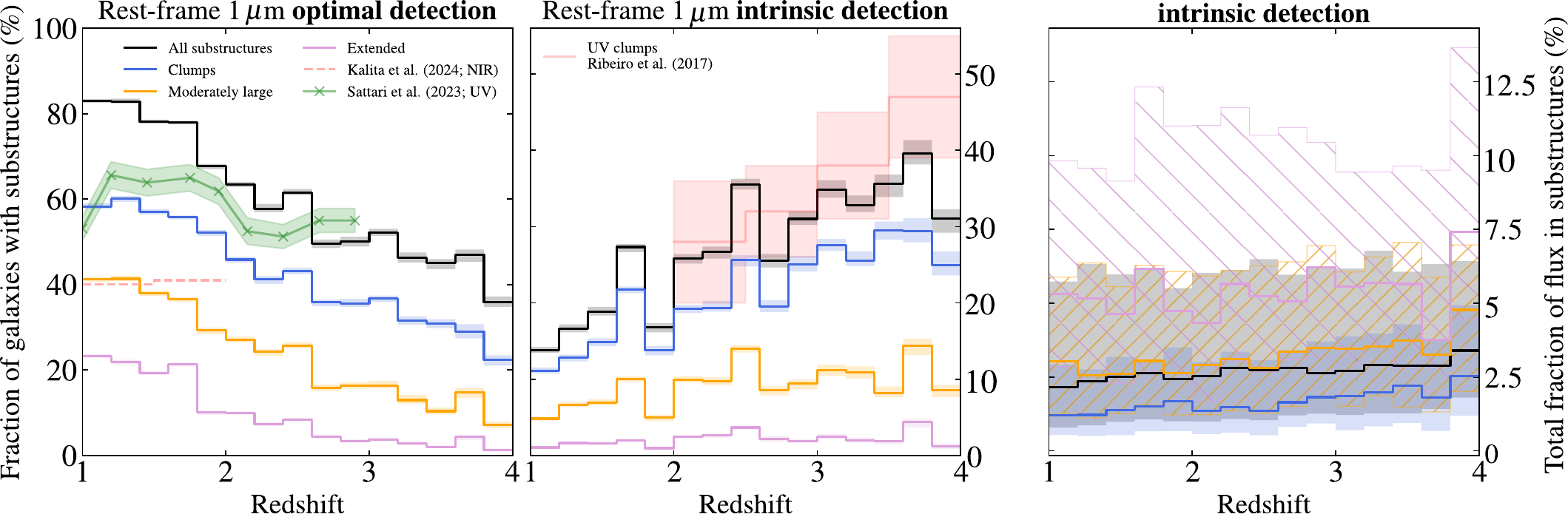}
    \caption{Detection of substructures at a rest-frame wavelength of \SI{1}{\micro\meter} for the entire sample (black) and for three substructure types: clumps (blue), moderately large ones (orange), and extended ones (pink). See Sect.\,\ref{sec:Results/size separation} for more details. \textbf{Left:} fraction of galaxies with at least one substructure detected with the optimal approach (see Sect.\,\ref{sec:detection/maximal detection}) as a function of redshift. We show the rest-frame NIR detections carried out by \citet{Kalita2024a} (red dashed line) and those in the rest-frame UV by \citet{Sattari2023} (green line and crosses). \textbf{Middle:} same but for the intrinsic detection approach (see Sect.\,\ref{sec:detection/intrisic detection}). We show as a light red line a comparable $2\sigma$ detection ($k_p = 2$, see reference) carried out in the rest-frame UV from \citet{Ribeiro2017}. \textbf{Right:} median fraction of flux in substructures (shaded and dashed areas correspond to the 16th and 84th percentiles of the distributions) as a function of redshift for galaxies with at least one substructure.}
    \label{fig:clump_fraction}
\end{figure*}

In this section, we consider the detection of substructures as a function of redshift using the optimal and intrinsic approaches. We consider all kind of substructures, whatever their size, and look at the fraction of flux found in substructures. We provide in Table\,\ref{table:fraction of substructures} the fraction of galaxies with substructures detected with the optimal and intrinsic detection approaches in various redshift bins.

\subsubsection{Optimal detection}
\label{sec:Results/overall/optimal}

To begin with, we consider the optimal detection approach. We show as a black line in the left panel of Fig.\ref{fig:clump_fraction} the fraction of galaxies with at least one substructure as a function of redshift. We find that we can detect substructures in more than 80\% of galaxies at $z = 1$, illustrating how ubiquitous they are. This fraction then smoothly decreases to about 40\% at $z = 4$. This is likely due to fainter substructures not observable at higher redshift because of surface brightness dimming (see Sect.\,\ref{sec:Results/overall/intrinsic}).
We compare our fractions to those measured in the rest-frame NIR by \citet{Kalita2024a} and in the rest-frame UV by \citet{Sattari2023}. Both works selected star-forming galaxies and do not correct for the effects of cosmological dimming and the angular diameter distance. Therefore, it is appropriate to compare them to the optimal detection approach. Galaxies in \citet{Kalita2024a} are at $1 < z < 2$ with $M_\star > \SI{e10}{\Msun}$, while for \citet{Sattari2023}, they are at $0.5 < z < 3$ with $M_\star > 10^{9.5}\,\si{\Msun}$.

In the redshift range $1 < z < 2$, we find that our fractions are systematically higher by a factor of two compared to those presented in Fig\,4 of \citet[][see also their Table\,1]{Kalita2024a}, with between approximately 65\% and 80\% of galaxies with at least one substructure in this redshift range compared to 40\% in their case.
This may be due to different sample selection and detection methods, in particular their use of a $5\sigma$ flux detection threshold which is higher than our value of $2\sigma$. Compared to the rest-frame UV detection of \citet{Sattari2023}, our detections in the rest-frame NIR are higher by about 15 percentile points at $z < 1.5$ but become compatible at higher redshift. Moreover, they find a decrease of the fraction of clumpy galaxies with increasing redshift which is consistent with our results, though our decrease is steeper than theirs. Given that they solely detect clumps and that we consider all types of substructures, this may be due to the stronger role that larger substructures play in our detections at $z \approx 1$, as illustrated by the pink and orange lines on the left-hand side of Fig.\,\ref{fig:clump_fraction}.

\subsubsection{Intrinsic detection}
\label{sec:Results/overall/intrinsic}

We show with a black line in the middle panel of Fig.\ref{fig:clump_fraction} the fraction of galaxies with at least one substructure detected using the intrinsic detection approach. At $z = 1$, we find a fraction of approximately 15\%, well below the 80\% found with the optimal detection technique. This is due to the fact that this approach forces the detections to be carried out as if the galaxies were at $z = 4$, thus discarding faint substructures not accessible at higher redshift. Hence, this dramatic drop from 80\% to 15\% at $z = 1$ between the two approaches hints at how frequently faint substructures are likely to be missed at high redshift.

From $z = 1$ to $z = 4$, we find that the fraction of galaxies with substructures in the rest-frame NIR increases from about 15\% to 30\%. This is broadly consistent with \citet{Ribeiro2017} who computed a similar fraction in the rest-frame UV with \HST{} for a sample of star-forming galaxies at $2 < z < 4$. However, we note that their approach has key differences with ours. First, they detect substructures that can span an area roughly twice as small. Second, they do not take into account the effect of the angular diameter distance on the size of their substructures. However, this effect is small compared to surface brightness dimming. And lastly, they effectively detect substructures at a flux level of $2\sigma$ at $z = 2$ whereas our approach detects them at a similar level at $z = 4$. Despite such differences, in what follow, \citet{Ribeiro2017} remains our main frame of comparison between the rest-frame UV and NIR given their similar treatment of surface brightness dimming.

These results indicate that substructures in the rest-frame NIR are common in galaxies at $1 < z < 4$, similarly to what is observed with \HST{} in the rest-frame UV.
We show in the right-most panel of Fig.\,\ref{fig:clump_fraction} the median fraction of rest-frame NIR flux located in substructures detected with the intrinsic approach as a function of redshift\footnote{For each galaxy we sum the flux of all substructures and divide by the total flux.}. To compute the fraction, we consider galaxies that have at least one substructure detected with the intrinsic approach, otherwise the median fraction would be null since more than half of the galaxies do not host any substructure.
Overall, substructures account for 3\% of the NIR flux, exhibiting minimal evolution with redshift, which is significantly less than what is seen in the rest-frame UV. If we used the optimal detection approach instead, the flux fraction would be increased by at most two percentile points, particularly at $z \approx 1$. For comparison, in \citet{Guo2015} and for galaxies at $2 < z < 3$, they measure an average clump contribution to the rest-frame UV light between 15\% at $M_\star \sim 10^{9.5}\,\si{\Msun}$ and 25\% at $10^{10.5}\si{\Msun}$. Furthermore, they measure a mild increase with decreasing redshift at the low-mass end with approximately 20\% at $M_\star \sim 10^{9.5}\,\si{\Msun}$ and $1 < z < 2$. 
We note the following two points. First, they measure the contribution of clumps to the rest-frame UV by including galaxies that do not host any clump. 
Second, they estimate their rest-frame UV flux fraction with clumps whose flux contribution is higher than 8\% in order to exclude faint clumps which they find to resemble local star-forming regions that they want to exclude from their analysis.
Therefore, compared to our flux fractions, their values are underestimated and the difference between the fractions in the rest-frame UV and NIR is, in practice, likely larger than what is quoted above.
In conclusion, while NIR substructures appear as ubiquitous as UV clumps, their contribution to the galaxies' light seems much weaker.

\begin{table*}
\caption{Fraction of galaxies with at least one substructure detected at a rest-frame wavelength of \SI{1}{\micro\meter} as a function of redshift.}
\centering
\begin{tabular}{ccccccccccccccc}

\hline\noalign{\vskip 2pt}
Redshift interval & $[1.0, 1.2)$ & $[1.2, 1.4)$ & $[1.4, 1.6)$ & $[1.6, 1.8)$ & $[1.8, 2.0)$ & $[2.0, 2.2)$ & $[2.2, 2.4)$ & $[2.4, 2.6)$  &\\
\noalign{\vskip 2pt}
Intrinsic - All & 13\% & 16\% & 18\% & 27\% & 16\% & 25\% & 26\% & 35\%\\
Optimal - All & 83\% & 82\% & 77\% & 77\% & 67\% & 63\% & 57\% & 61\%\\
Intrinsic - clumps & 13\% & 16\% & 18\% & 27\% & 16\% & 25\% & 26\% & 35\%\\
Optimal - clumps & 58\% & 60\% & 56\% & 55\% & 51\% & 45\% & 41\% & 43\%\\
\hline
\hline\noalign{\vskip 2pt}
Redshift interval & $[2.6, 2.8)$ & $[2.8, 3.0)$ & $[3.0, 3.2)$ & $[3.2, 3.4)$ & $[3.4, 3.6)$ & $[3.6, 3.8)$ & $[3.8, 4.0]$\\
\noalign{\vskip 2pt}
Intrinsic - All & 25\% & 30\% & 34\% & 33\% & 35\% & 39\% & 31\%\\
Optimal - All & 49\% & 49\% & 52\% & 46\% & 44\% & 46\% & 36\%\\
Intrinsic - clumps & 25\% & 30\% & 34\% & 33\% & 35\% & 39\% & 31\%\\
Optimal - clumps & 35\% & 35\% & 36\% & 31\% & 30\% & 28\% & 22\%\\[2pt]
\hline\noalign{\vskip 2pt}

\end{tabular}

\label{table:fraction of substructures}
{\small\raggedright {\bf Notes:} We separate between all kinds of detections and clumps only (see Sect.\,\ref{sec:Results/size separation}).\par}
\end{table*}

\subsection{Separating substructures in size}
\label{sec:Results/size separation}

In this section, we consider the detection of different types of substructures as a function of redshift for the optimal and intrinsic detection approaches. We split our sample into three size categories that are clumps, moderately large substructures, and extended ones (see Sect.\,\ref{sec:detection/sample cleaning} for their definitions). We provide in Table\,\ref{table:fraction of substructures} the fraction of galaxies with clumps detected with the optimal and intrinsic detection approaches in various redshift bins.

\subsubsection{Optimal detection}

We present in the left panel of Fig.\,\ref{fig:clump_fraction} the fraction of galaxies for each type of substructure detected with the optimal detection technique (clumps in blue, moderately large substructures in orange, extended ones in pink). 
For each class, we recover a trend similar to Sect.\,\ref{sec:Results/overall/optimal} whereby the fraction of galaxies with a given type of substructure systematically decreases with increasing redshift.
Moreover, we find that clumps are much more common (about 60\% at $z = 1$ and 20\% at $z = 4$) than moderately large substructures (about 40\% at $z = 1$ and 8\% at $z = 4$) that are themselves more common than extended substructures (about 23\% at $z = 1$ and 1\% at $z = 4$). 
By $z \approx 3$, very few extended substructure are detected in the optimal detection approach.
We note that these quite high fractions of galaxies with clumps are inconsistent with certain simulations \citep[e.g. \NIHAO{}, see][]{Buck2017} which do not produce substructures in stellar mass maps, but generate many in the UV maps. However, the presence of NIR clumps strongly varies between simulations based on the recipes used to, for instance, model feedback processes, the resolution of the simulations, and the fraction of gas within galaxies \citep[e.g.][]{Perez2013, Bournaud2014, Fensch2021}.

\subsubsection{Intrinsic detection}
\label{sec:Results/size separation/intrinsic}

For the intrinsic detection approach, trends are different. This is shown in the middle panel of Fig.\,\ref{fig:clump_fraction}. The detection of moderately large substructures marginally increases with redshift from roughly 5\% at $z = 1$ to 10\% at $z = 4$, while that of extended ones oscillates around $1\%$. Only the clump population shows an evolution with redshift that is consistent with the total population, increasing from 10\% at $z = 1$ to $25\%$ at $z = 4$. This is expected since clumps dominate the detection budget (see Fig.\,\ref{fig:histogram area of substructures}). Yet, the contribution of moderately large substructures is non-negligible, especially at low redshift.
The rise of rest-frame NIR clump detections with increasing redshift is consistent with the results found by \citet{Ribeiro2017} in the rest-frame UV. Still, it is worth noticing that our detections are systematically lower and that our redshift evolution is shallower than theirs. Indeed, they find a rest-frame UV clump fraction of about 27\% at $z = 1$ (a factor of 1.3 above our NIR fraction) and of 57\% at $z = 4$ (a factor of 2.3 above). Similarly to \citet{Kalita2024a}, we do not share a consensus with \citet{Ribeiro2017} on the exact definition of what we call clumps and, therefore, they include in their sample larger substructures than we do, as is illustrated in their Fig.\,10. This certainly explains why our fractions in the rest-frame NIR for all types of substructures are closer to theirs in the rest-frame UV than when we are considering clumps alone.

We show in the right panel of Fig.\,\ref{fig:clump_fraction} the fraction of flux found in each class of substructures detected with the intrinsic approach. Here, for a given class (clumps, moderately large, or extended), we consider galaxies that have at least one substructure in that class. 
As expected, extended substructures contain the largest fraction of flux in the NIR with typical values ranging from 3\% to 12\%, followed by moderately large ones (between 2\% and 7.5\%), and then clumps (between 1\% and 5\%). 
Furthermore, we do not find any redshift evolution for extended and moderately large substructures. While we do find a mild trend for clumps, with the median NIR flux fraction increasing from around 1\% at $z = 1$ to 2.5\% at $z = 4$, it is mostly encompassed within the scatter of the order of 2\%. 
We note that clumps nevertheless dominate the substructure detection budget (as illustrated by the intrinsic detections in Fig.\,\ref{fig:clump_fraction}). Therefore, even though extended substructures are notably brighter when present, their prevalence is so low that they contribute little to the flux budget in general. In other terms, 
\begin{enumerate*}
    \item NIR clumps appear nearly as ubiquitous as UV clumps,
    \item their detection in galaxies similarly increases with increasing redshift, but
    \item their NIR flux is so low that their contribution to the flux budget of their host galaxy is negligible ($\sim 1.5\%$). This is in sharp contrast with UV clumps that can typically contribute 30\% to the UV flux of their host galaxy \citep[e.g.][]{Guo2015, Guo2018, Ribeiro2017}. We note that, if these two clump populations are similar, this seems consistent with clumps that are relatively long-lived, though still with a strong UV component, observed in simulations of isolated gas-rich disk galaxies at $z \approx 2$ \citep[e.g.][]{Bournaud2014}.
\end{enumerate*}

Finally, it is interesting to compare the results for the optimal and intrinsic detections since the optimal approach is supposed to be more efficient at detecting faint substructures. At $z = 1$, the detection of clumps is about 60\% in the optimal case but 10\% in the intrinsic case. For extended substructures, it is approximately 23\% in the optimal case and 1\% in the intrinsic one. This highlights that a large number of faint, and thus probably low-mass, clumps are missed by the intrinsic approach at low redshift.
Assuming that a similar population of fainter clumps exists at higher redshift, it shows that our detections are biased to the brightest clumps and that galaxies are certainly even clumpier than they appear. This interpretation is consistent with lensed observations of clumps in the optical and the NIR \citep[e.g.][]{claeyssens_star_2023, Claeyssens2024}.

\subsection{Detecting substructures across the \texorpdfstring{$M_\star - {\rm SFR}$}{Mstar-SFR} plane}
\label{sec:Results/MS}

In this section, we only consider substructures detected with the intrinsic detection approach.
Historically, UV clumps have been studied in star-forming galaxies on the MS and in starbursts \citep[e.g.][]{Wuyts2012, Guo2015, Guo2018, Ribeiro2017, Sattari2023}. Several studies have considered the effect of the host galaxy's stellar mass on clump statistics \citep[e.g.][]{Guo2015, Guo2018} and, indirectly, of the SFR by measuring the fractional UV luminosity in clumps. This also applies to NIR clumps \citep{Kalita2024a, Kalita2024}, though the relation to the host's SFR has been little studied (except for the fractional SFR; see Fig.\,11 of \citealt{Kalita2025}). 
Furthermore, regarding extended substructures, a few studies have measured the strength of spiral arms as a function of the stellar mass and SFR of the host galaxies \citep[e.g.][]{Yu2021}.
In this section, we look at the interplay between the host's stellar mass and SFR on the presence and properties of NIR substructures.

\subsubsection{Fraction of galaxies with substructures}
\label{sec:Results/MS/fraction}

\begin{figure*}[htbp]
    \centering
    \includegraphics[width=\linewidth]{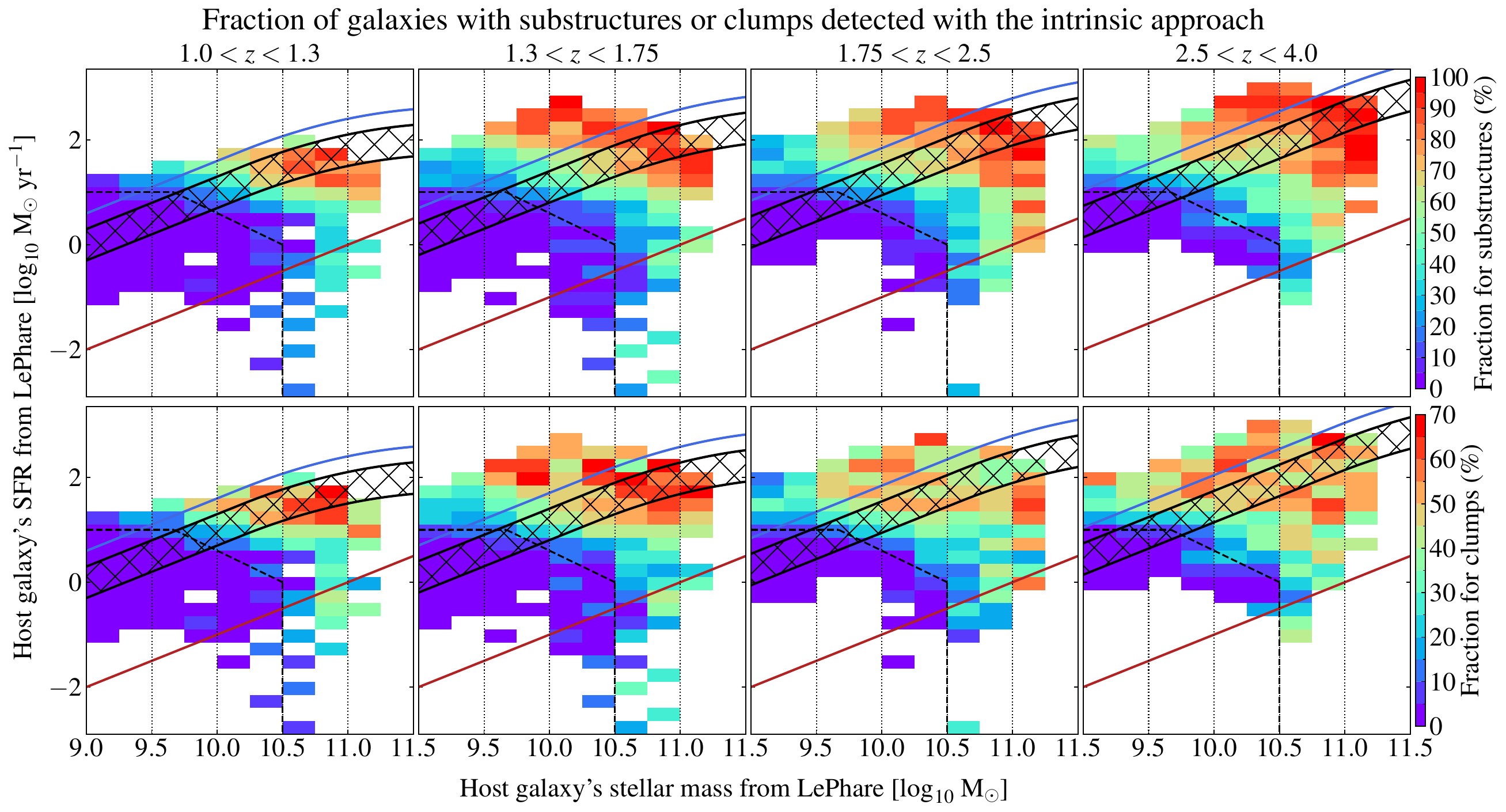}
    \caption{\textbf{Top row:} median fraction of galaxies with at least one substructure detected with the intrinsic approach as a function of their host galaxy's stellar mass and SFR in four redshift bins, each spanning roughly \SI{1}{\giga\year} of cosmic time. \textbf{Bottom row:} same for clumps. Violet corresponds to 0\% and red to 100\% for substructures (70\% for clumps). In each panel, the dashed line isolates the region where no substructure is detected (see Eq.\,\ref{eq:delimiting line}). Galaxies above the blue line are located in the starburst region, those in the hatched region are on the MS from \citet{Schreiber2015} evaluated at the average redshift of each panel, and those below the red line are in the red sequence with ${\rm sSFR} < \SI{e-11}{\per\year}$. We do not show bins with fewer than ten galaxies.}
    \label{fig:fraction_vs_MS}
\end{figure*}

According to \citet{Arango2024}, we identify the MS by selecting galaxies that have a SFR within $\pm \SI{0.3}{\dex}$ in the relation from \citet{Schreiber2015}. Furthermore, we adopt similar definitions to select starbursts ($\SI{0.6}{\dex}$ above the MS), the red sequence (${\rm sSFR} < \SI{e-11}{\per\year}$), and the green valley (between the MS and the red sequence). We show in Fig.\,\ref{fig:fraction_vs_MS} the fraction of galaxies with at least one rest-frame NIR substructure (top row) and one clump (bottom row) detected with the intrinsic detection technique in bins of host galaxy's stellar mass and SFR. We split our sample into four redshift bins from $z = 1$ to $z = 4$, with each step chosen so that it encompasses \SI{1}{\giga\year} of cosmic history. This choice seems a good compromise for three reasons.
First, it provides a sufficiently large number of galaxies in each redshift bin to probe different parts of the $M_\star - {\rm SFR}$ plane. Second, the redshift steps are small enough to follow the evolution of the galaxy population and its clumpiness before and after the peak of star-formation at $z \approx 2$. And third, it corresponds to the typical migration timescale of long-lived clumps produced in certain simulations \citep[e.g.][]{Bournaud2011, Dubois2012}.
Furthermore, we provide in Fig.\,\ref{fig:substructures vs MS with Cigale} a similar figure using the stellar mass and SFR derived with \Cigale{}. Overall, even though the stellar masses from \Cigale{} are above those of \LEPHARE{} and the SFRs slightly below, we recover similar results to those presented below. 

First, along the MS and at all redshifts, we observe an increasing number of galaxies with substructures as we move towards higher host's stellar mass and SFR values. Below $10^{9.75}\,\si{\Msun}$ at $z < 1.3$ and $10^{9.5}\,\si{\Msun}$ at $z > 1.3$, we do not find substructures on the MS. Above $10^{10.5}\,\si{\Msun}$, the fraction of MS galaxies with substructures becomes greater than 70\% at all redshifts. 
Beyond $10^{10.75}\,\si{\Msun}$, nearly all galaxies on the MS have at least one substructure. 
We find a similar trend for clumps at $z < 1.75$, with a fraction of galaxies with clumps reaching about 50\% at $10^{10.5}\,\si{\Msun}$ and 70\% at $M_\star > 10^{10.75}\,\si{\Msun}$. However, the highest fraction of galaxies with clumps ($\sim 70\%$) is reached at the high-mass end of the MS at $1.3 < z < 1.75$ and then reduces to approximately 50\% at $z > 1.75$.
Second, there is a noticeable increase of substructure detections (including clumps) with higher SFRs at a fixed stellar mass, with the highest detection rates reached for starbursts and the high-mass end of the MS. 
This result appears consistent with measurements of the strength of spiral arms measured in local galaxies selected in the Sloan digital sky survey (SDSS) where stronger arms are found in more star-forming galaxies at a given stellar mass \citep[see Fig.\,2 for][]{Yu2021}.
Overall, we find higher fractions of galaxies with substructures or clumps at higher stellar masses at a given SFR. Interestingly, this applies for galaxies in the green valley and in the red sequence with, for the latter, the detections ramping up from nearly 0\% at the lowest stellar masses to about 40\% at $M_\star > 10^{10.5}\,\si{\Msun}$. Similarly, the fraction of galaxies on the green valley with at least one substructure increases from 0\% at $M_\star \lesssim \SI{e10}{\Msun}$ to 100\% at $M_\star \approx \SI{e11}{\Msun}$. Furthermore, the galaxies at the high-mass end of the green valley behave differently depending on their SFR. Those with the highest SFRs have fractions close to those found on the MS at the same mass. On the other hand, massive galaxies on the green valley with the lowest SFRs have fractions of roughly 50\% at $z < 2.5$ and 60\% at $z > 2.5$, closer to the values found at the high-mass end of the red sequence. 

According to Fig.\,\ref{fig:fraction_vs_MS}, the region where we do not detect any substructures appears mostly independent of redshift. We can define a delimiting line that isolates that region, as illustrated with the dashed line in Fig.\,\ref{fig:fraction_vs_MS} whose parametric form is given by:

\begin{equation}
\begin{cases}
    y = 10 &\mbox{for} \quad 9 < x < 9.67, \\
    y = -1.2 x + 12.6 &\mbox{for} \quad  9.67 < x < 10.5, \\
    x = 10.5 & \mbox{for} \quad  y < 0,
\end{cases}
\label{eq:delimiting line}
\end{equation}
with $x = \log_{10} M_\star$ in \si{\Msun} and $y = \log_{10} \rm SFR$ in \si{\Msun\per\year}. We estimate the uncertainty associated with each bin using bootstrapping (see Sect.\,\ref{sec:Introduction}) and find that the standard deviation of the fraction of galaxies with substructures is of the order of 10 percentile points. Galaxies in the region delimited by Eq.\,\ref{eq:delimiting line} tend to have uncertainties below 2 percentile points. On the other hand, low-mass starbursts as well as galaxies in the green valley and red sequence with $M_\star \gtrsim 10^{10.5}\,\si{\Msun}$ have a higher uncertainty of 15 percentiles points, with a few bins reaching 25 percentiles points.

It is expected from the mass-size relation \citep[e.g.][]{Martorano2024} that the size and compactness of the host galaxies will depend on their stellar mass and SFR. Lower-mass galaxies and/or with lower SFR should be smaller, thus reducing the area over which clumps are detected (as illustrated in Fig.\,\ref{fig:detection area vs radius}) and may disfavor small galaxies compared to extended ones. 
We have checked for this effect by reproducing Fig.\,\ref{fig:fraction_vs_MS} in five bins of global effective radius (see Eq.\,6 of \citealt{Mercier2022}): $R_{\rm eff} < \SI{1.08}{\kilo\parsec}$, $1.08 < R_{\rm eff} / \si{\kilo\parsec} < 1.61$, $1.61 < R_{\rm eff} / \si{\kilo\parsec} < 2.22$, $2.22 < R_{\rm eff} / \si{\kilo\parsec} < 3.17$, and $R_{\rm eff} > \SI{3.17}{\kilo\parsec}$. These bins are chosen so that they contain 20\% of the galaxies.
We find that, whatever the size bin, we always recover Eq.\,\ref{eq:delimiting line} meaning that the part of the $M_\star - \rm SFR$ plane that is devoid of substructures is not biased by the size of the galaxies. 
The high-mass end of the red sequence ($M_\star > 10^{10.5}\,\si{\Msun}$) is dominated by small ($R_{\rm eff} < \SI{1.61}{\kilo\parsec}$) and therefore compact galaxies, which goes against a size bias. 
However, the variation of the fraction of galaxies with substructures or clumps visible along the MS and the starburst region seen at $M_\star \gtrsim \SI{e10}{\Msun}$ in Fig.\,\ref{fig:fraction_vs_MS} becomes prominent at $R_{\rm eff} > \SI{1.61}{\kilo\parsec}$ and is dominated by large galaxies ($R_{\rm eff} > \SI{2.2}{\kilo\parsec}$). Furthermore, massive galaxies in the green valley behave similarly to those on the MS at the same mass.
Therefore, this may bias our substructure and clump detections at the high-mass and high-SFR ends by favoring the most massive galaxies with the highest SFRs. To control for this effect, we reproduce Fig.\,\ref{fig:fraction_vs_MS} but weighting each substructure detection by the area over which it is carried out, as illustrated in Fig.\,\ref{fig:substructures vs MS with area weigths}. We find that we recover comparable results as when not weighting the detections, which shows that our results are not produced by a galaxy size bias but appear as a genuine stellar mass and SFR dependence.
We note that the lack of detections in the region delimited by Eq.\,\ref{eq:delimiting line} is likely not intrinsic but an observational effect. If we compare our detections in the lowest redshift bin using the intrinsic and optimal detection techniques, we see that the delimiting line given by Eq.\,\ref{eq:delimiting line} moves towards galaxy hosts with lower stellar mass and SFR values. In other terms, low-mass and low-SFR galaxies are likely to host substructures that are too faint to be detectable with the intrinsic detection method. We do not consider such substructures in what follows because we cannot observe them at higher redshift.

\subsubsection{Flux in substructures}
\label{sec:Results/MS/flux}

\begin{table}[]
    \centering
    \caption{Correlation between the fraction of rest-frame NIR flux in substructures or clumps and sSFR and partial correlation between flux fraction and (i) stellar mass of the host galaxy at fixed host's SFR, and (ii) SFR of the host galaxy at fixed host's stellar mass.}
    \resizebox{\linewidth}{!}{
    \begin{tabular}{lccc}
        \hline\noalign{\vskip 2pt}
        \multirow{2}{*}{Redshift interval}& \multicolumn{2}{c}{Partial correlation with} & Correlation with\\
         & $\log_{10} M_\star$ & $\log_{10} \rm SFR$ & $\log_{10} \rm sSFR$\\
        \hline
        \hline\noalign{\vskip 2pt}
        \multicolumn{4}{c}{\bf Substructures} \\[2pt]
        $[1.0\hphantom{0}, 1.3)$ & $-0.15^{+0.05}_{-0.05}$ & $0.07^{+0.05}_{-0.06}$ & $0.15^{+0.05}_{-0.06}$ \\[2pt]
        $[1.3\hphantom{0}, 1.75)$ & $-0.07^{+0.04}_{-0.06}$ & $0.15^{+0.03}_{-0.04}$ & $0.17^{+0.03}_{-0.04}$ \\[2pt]
        $[1.75, 2.5)$ & $-0.05^{+0.03}_{-0.03}$ & $0.16^{+0.03}_{-0.03}$ & $0.17^{+0.03}_{-0.03}$\\[2pt]
        $[2.5\hphantom{0}, 4.0]$ & $-0.13^{+0.04}_{-0.02}$ & $0.14^{+0.03}_{-0.03}$ & $0.17^{+0.03}_{-0.03}$\\[2pt]
        \hline\noalign{\vskip 2pt}
        \multicolumn{4}{c}{\bf Clumps} \\[2pt]
        $[1.0\hphantom{0}, 1.3)$ & $-0.54^{+0.05}_{-0.04}$ & $-0.08^{+0.06}_{-0.06}$ & $0.23^{+0.05}_{-0.06}$\\[2pt]
        $[1.3\hphantom{0}, 1.75)$ & $-0.49^{+0.03}_{-0.03}$ & $-0.08^{+0.04}_{-0.04}$ & $0.23^{+0.04}_{-0.04}$\\[2pt]
        $[1.75, 2.5)$ & $-0.51^{+0.03}_{-0.03}$ & $-0.11^{+0.04}_{-0.03}$ & $0.24^{+0.03}_{-0.04}$ \\[2pt]
        $[2.5\hphantom{0}, 4.0]$ & $-0.47^{+0.03}_{-0.03}$ & $-0.10^{+0.04}_{-0.03}$ & $0.23^{+0.04}_{-0.03}$\\[2pt]
        \hline\noalign{\vskip 2pt}
    \end{tabular}}
    \label{tab:correlations flux with mass sfr}

    {\small\raggedright {\bf Notes:} Values are computed using the Pearson correlation and uncertainties correspond to the 95\% confidence interval. Using the Spearman correlation leads to similar trends but with stronger correlation coefficients, thus showing non-linearity. \par}
\end{table}

\begin{figure*}[htbp]
    \centering
    \includegraphics[width=\linewidth]{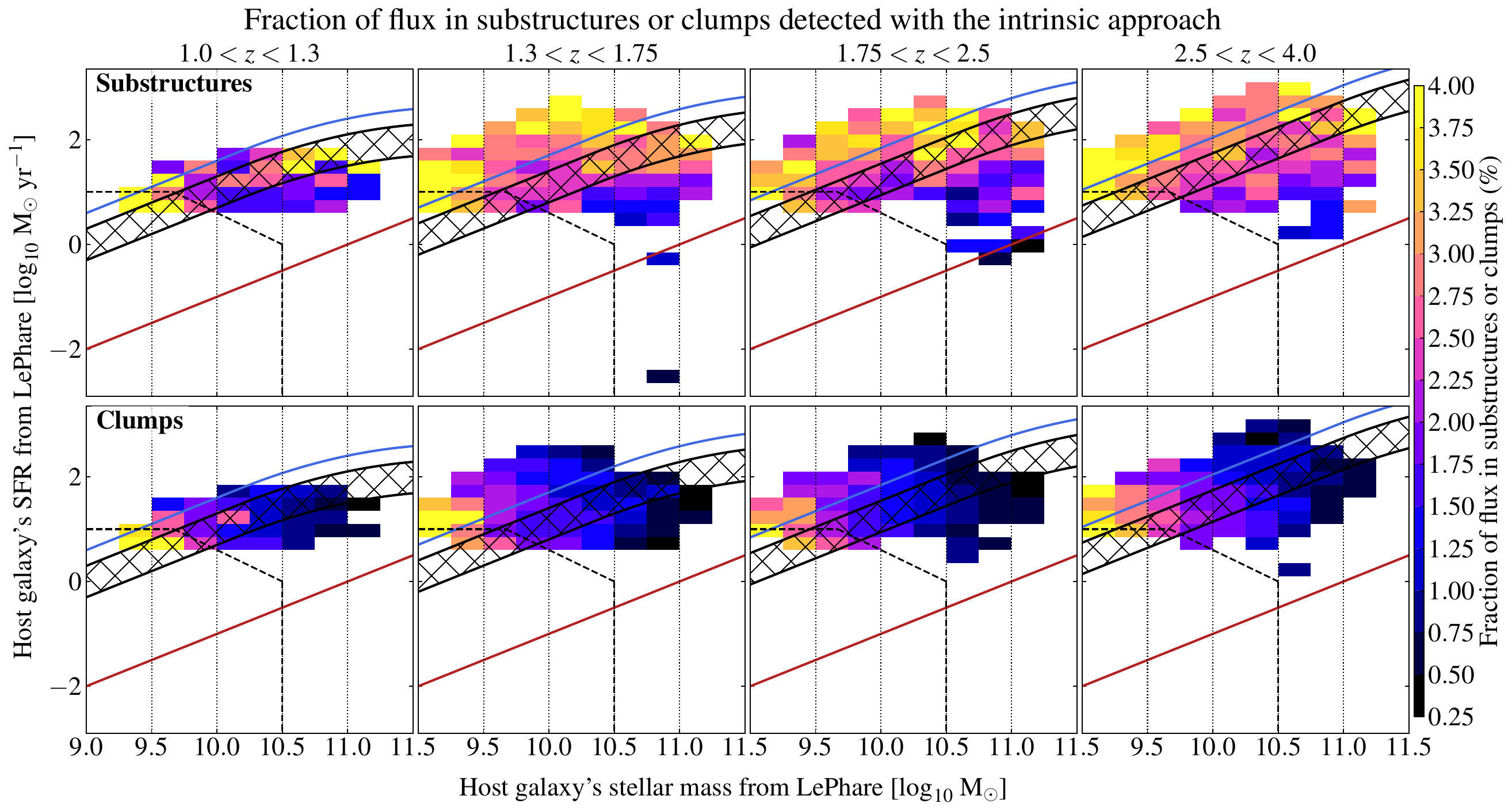}
    \caption{{\bf Top row:} same as Fig.\,\ref{fig:fraction_vs_MS} but representing in each galaxy host's stellar mass and SFR bin the fraction of flux in substructures detected using the intrinsic detection approach. To compute the fraction of flux we consider only those galaxies that contain at least one substructure. {\bf Bottom row:} same for clumps.}
    \label{fig:flux_fraction_vs_MS}
\end{figure*}

We consider the total fraction of rest-frame NIR flux in substructures and clumps detected with the intrinsic detection approach across the $M_\star - {\rm SFR}$ plane. This is shown in Fig.\,\ref{fig:flux_fraction_vs_MS} for galaxies that host at least one substructure (top row) or one clump (bottom row) and for stellar mass and SFR bins containing at least ten galaxies. Therefore, Fig.\,\ref{fig:flux_fraction_vs_MS} is limited to galaxies on the MS, in the starburst region, and at the high-mass and high-SFR ends of the green valley.
Alternatively, we also show in Fig.\,\ref{fig:flux fraction vs mass with SFR as color} the median fraction of rest-frame NIR flux in substructures or clumps as a function of the host galaxy's stellar mass, color coded in bins of host galaxy's SFR. The difference between Figs.\,\ref{fig:flux_fraction_vs_MS} and \ref{fig:flux fraction vs mass with SFR as color} is that the latter also highlights the scatter around the median value for populations of galaxies with different star formation levels.

The brightest substructures at $z > 1.3$ are found in the starburst region throughout the entire stellar mass range. They typically host 4\% of the total rest-frame NIR flux of the galaxies, with fractions oscillating between approximately 2\% and 10\% at $M_\star \approx \SI{e9}{\Msun}$ and between 2\% and 7.5\% at higher stellar masses. 
From Fig.\,\ref{fig:flux_fraction_vs_MS}, it looks like stellar mass is loosely correlated with the flux fraction and that galaxies with a lower SFR have a lower fraction of their rest-frame NIR flux in substructures. Indeed, the lowest flux fraction, equal to 0.5\%, is reached at $ 1.75 < z < 2.5$ at the limit between the high-mass ends of the green valley and the red sequence.
We estimate the partial correlation between the rest-frame NIR flux fraction and
\begin{enumerate*}
    \item the galaxy host's stellar mass at fixed host's SFR and
    \item the host' SFR at fixed host's stellar mass.
\end{enumerate*}
In addition, we consider the correlation between the rest-frame NIR flux fraction and the sSFR of the host galaxies. The values of the (partial) correlations computed in four redshift bins are presented in Table\,\ref{tab:correlations flux with mass sfr}. Overall, we find low correlations between the rest-frame NIR flux fraction in substructures and the three aforementioned galaxy host's physical parameters. At $1 < z < 1.3$, the flux fraction is loosely anti-correlated with stellar mass (Pearson coefficient of $r \approx - 0.15$) but, at higher redshift, the correlation vanishes and the flux fraction becomes loosely correlated with SFR ($r \approx 0.15$) instead. At any redshift, there is also a weak correlation with sSFR ($r \approx 0.15$). This means that more star-forming galaxies are likely to contain more of their rest-frame NIR flux in substructures. This is consistent with starbursts having the highest fractions, as previously mentioned.

Interestingly, the trends found for clumps are different, with the highest rest-frame NIR median flux fraction (about 4\%) reached in the low-mass part of the MS and starburst region ($M_\star < 10^{9.5}\,\si{\Msun}$). The fraction decreases with increasing galaxy host's stellar mass and SFR, reaching a median value of less than 0.5\% at $M_\star \sim \SI{e11}{\Msun}$ on the MS, starburst region, and green valley.
Similarly to substructures, we derive the correlation with the host galaxies' sSFR and the partial correlations with stellar mass and SFR. We find a negligible anti-correlation between the median clump flux fraction and the galaxy host's SFR but a much stronger anti-correlation with the stellar mass ($r \approx - 0.5$). There is also a weaker positive correlation with the host's sSFR ($r \approx 0.23$) but the stellar mass remains the physical parameter that is the most strongly correlated. This is consistent with clumps in massive galaxies contributing less to the rest-frame NIR flux budget than in lower-mass galaxies.
We note that this corresponds to a mild evolution of the stellar mass found in clumps with the galaxy host's stellar mass. Assuming similar mass-to-light ratios between the clumps and the host galaxies, a flux fraction of 4\% at $M_\star = 10^{9.5}\,\si{\Msun}$ corresponds to a total clump stellar mass of roughly \SI{e8}{\Msun}, while 1\% at $M_\star = \SI{e11}{\Msun}$ corresponds to a total clump stellar mass of roughly \SI{5e8}{\Msun}.

If we use the non-parametric Spearman correlation instead, we find the same trends for both clumps and substructures with, generally speaking, correlations stronger by 20\%. This should be taken as an indication that the fraction of flux in substructures and clumps in the rest-frame NIR is not linearly correlated to the galaxy host's stellar mass, SFR, and sSFR. Furthermore, we note that, in either case, we do not find an evolution of the correlations with redshift.
The different correlations with the stellar mass and SFR of the host galaxy observed between all types of substructures and clumps show that the flux contribution of moderately large and extended substructures behave differently from clumps and are more strongly correlated with SFR and sSFR than with stellar mass. We note that this behavior appears consistent with the results from \citet{Yu2021} where they find stronger spiral arms in galaxies with higher SFR values, but no dependence on the galaxy host's stellar mass.

\section{Discussion}
\label{sec:discussion}

\subsection{Interpretation about the evolution of clumpiness with redshift}
\label{sec:discussion/GV and RS}

\begin{figure*}
    \centering
    \includegraphics[width=\linewidth]{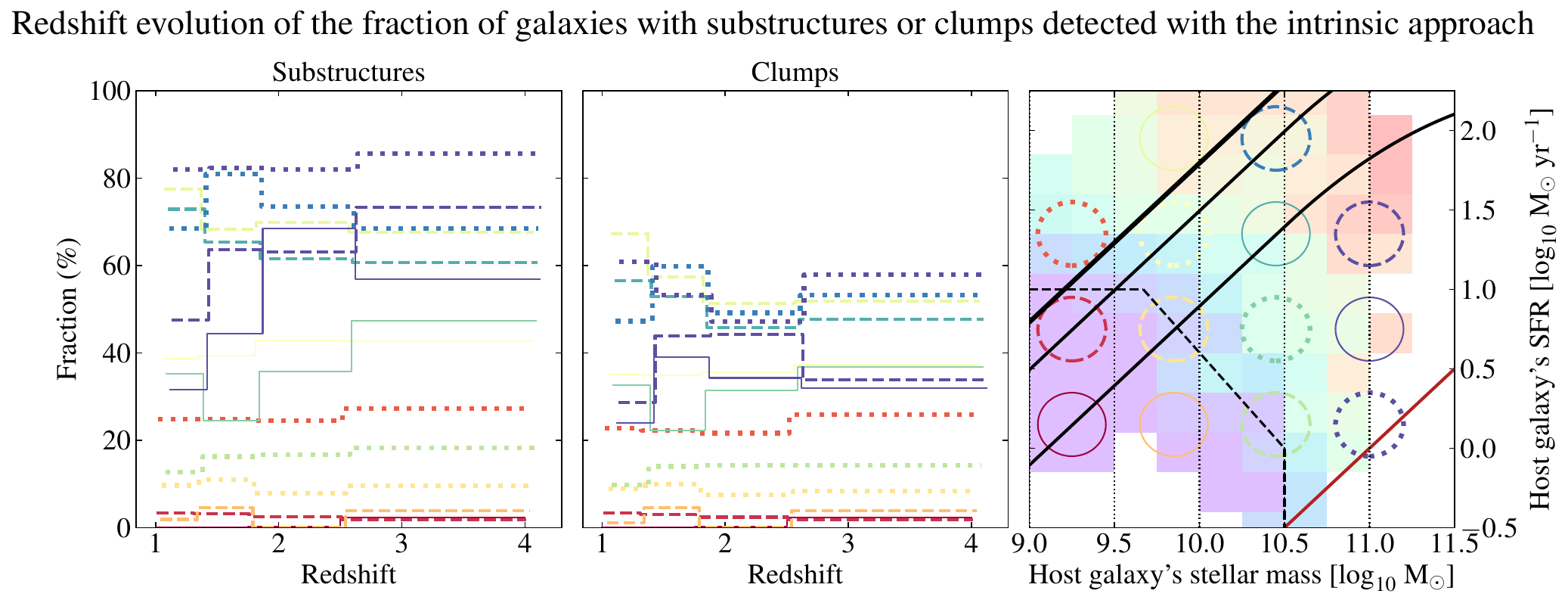}
    \caption{Evolution of the fraction of galaxies with substructures (left panel) and clumps (middle panel) detected with the intrinsic detection approach as a function of redshift for different regions of the $M_\star - {\rm SFR}$ plane, as illustrated in the right panel with circles of different color, line style, and thickness. We use the same redshift bins spanning \SI{1}{\giga\year} of cosmic time as in Figs.\,\ref{fig:fraction_vs_MS} and \ref{fig:flux_fraction_vs_MS}. Each circular region has a radius of \SI{0.2}{\dex} and the regions are chosen to evenly map the plane. We only show regions with more than ten galaxies in all redshift bins.}
    \label{fig:std of MS}
\end{figure*}

We show in Fig.\,\ref{fig:std of MS} the redshift evolution of the fraction of galaxies with substructures (left panel) and clumps (middle panel) in different regions of the $M_\star - {\rm SFR}$ plane, as illustrated in the right panel.
Overall, we do not see any noticeable trend such that the fraction of galaxies with clumps or substructures at a given location in the $M_\star - {\rm SFR}$ plane would increase or decrease with redshift. The observed fluctuations from one redshift bin to the next of, typically, ten percentile points are within the uncertainties.
There are, however, two regions at the high-mass end of the green valley ($M_\star = \SI{e11}{\Msun}$) with $\log_{10} ({\rm SFR} / \si{\Msun\per\year}) = 0.15$ and 0.75 that show a measurable decrease with redshift going from approximately 60\% and 75\% at $z = 4$ to 50\% and 35\% at $z = 1$, respectively.
These results may seem contradictory to the redshift evolution found in Sects.\,\ref{sec:Results/overall} and \ref{sec:Results/size separation} but it can be explained in two complementary ways.
First, as cosmic time progresses, we see the buildup of the green valley and the red sequence which both have lower fractions of galaxies with substructures than the MS or starburst region. This is particularly true in the low-mass regime ($M_\star < 10^{10.5}\,\si{\Msun}$) where we do not detect any substructures.
Second, galaxies on the MS and starbursts evolve with cosmic time, both moving towards lower SFRs at a given stellar mass. Therefore, this leads overall to lower fractions of galaxies with substructures or clumps. 

\begin{figure}[htbp]
    \centering
    \includegraphics[width=\linewidth]{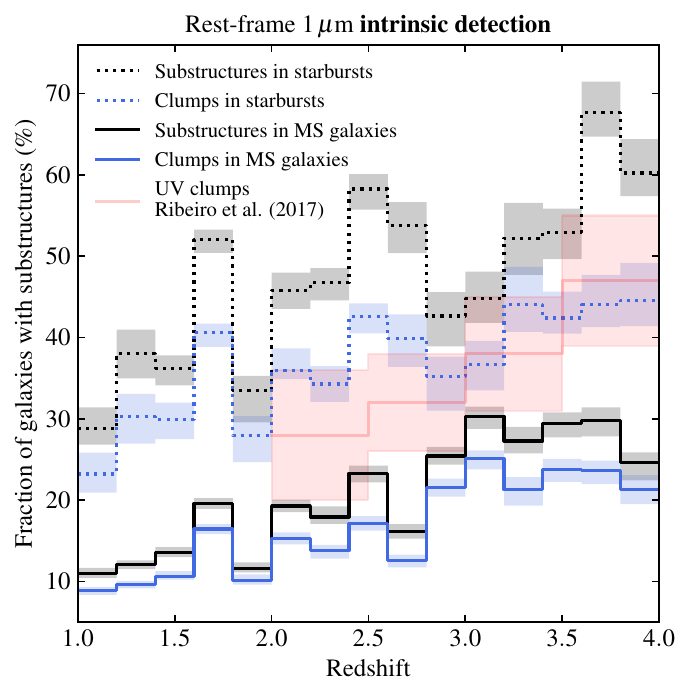}
    \caption{Intrinsic detection of substructures (black) and clumps (blue) for galaxies on the MS (continuous lines) and for starbursts (dotted lines) as a function of redshift. For comparison, we also show the rest-frame UV detections from \citet[][light red]{Ribeiro2017}.}
    \label{fig:fraction of clumps MS and SB}
\end{figure}

These two effects are illustrated in Fig.\,\ref{fig:fraction of clumps MS and SB} where we see that the fraction of clumpy galaxies on the MS and in the starburst region both increase at least by a factor two from $z = 1$ to $z = 4$.
In practice, the probability for a galaxy to host substructures will vary depending on its evolutionary track within the $M_\star - {\rm SFR}$ plane. For instance, a galaxy starting with low stellar mass and SFR values and building up its mass by moving towards the MS or the starburst region is more likely to host substructures at a later stage. On the other, a low-mass galaxy at $z = 4$ starting in the starburst region is likely to host substructures. If it builds its mass at a constant SFR and then quenches at $M_\star \lesssim 10^{10.5}\,\si{\Msun}$, we expect it to become smoother with cosmic time. On the other hand, if it quenches at $M_\star \gtrsim 10^{10.5}\,\si{\Msun}$, we do not expect all substructures to disappear, though, according to Fig.\,\ref{fig:flux_fraction_vs_MS}, they should become fainter with cosmic time.

\subsection{Impact of young stellar populations}
\label{sec:discussion/young populations}

The strong correlation between the clumpiness of the galaxy population and the SFR of the host galaxies raises the question of whether rest-frame NIR substructures in galaxies with high SFR values could be the result of young stellar populations. This could also explain the higher fractions of fluxes seen in Fig.\,\ref{fig:fraction_vs_MS}. We devise a toy model to check for this effect by assuming that substructures have been formed through a recent burst of star formation, so that the impact of young stellar populations in the rest-frame NIR is maximal. We can split the rest-frame NIR flux of the host galaxy into an old component $F_{\rm o} = M_\star / \Upsilon_\star$, with $\Upsilon_\star$ the mass-to-light ratio for the old stars, and a young one with $F_{\rm y} = \alpha_{\rm y} \times \rm SFR$ where $\alpha_{\rm y}$ is the SFR-to-flux conversion factor. The old component scales with $M_\star$ because its flux is produced by the accumulation of old stars from its past star formation history while the young component is short-lived and, thus, must scale with the SFR.
On the other hand, substructures are made of young low- and high-mass stars whose flux, by construction, must scale with the SFR as $F_{\rm s} = \alpha_{\rm y} \times C_{\rm SFR} \times  \rm SFR$, where $C_{\rm SFR}$ is the fraction of SFR found in substructures.
Let us note $\eta = F_{\rm s} / (F_{\rm o} + F_{\rm y})$ the fraction of rest-frame NIR flux in substructures, so that we have

\begin{equation}
    \eta = \frac{C_{\rm SFR}}{1 + \left ( \alpha_{\rm y} \Upsilon_\star \times \rm sSFR\right )^{-1}}.
    \label{eq:fraction of flux in NIR}
\end{equation}

If the fraction of SFR in substructures ($C_{\rm SFR}$) is constant, Eq.\,\ref{eq:fraction of flux in NIR} shows that $\eta$ should increase with the sSFR of the host galaxy. 
Similarly, if $C_{\rm SFR}$ increases with the sSFR (starbursts and galaxies in the MS are more likely to develop instabilities and produce clumpy star formation), we expect $\eta$ to increase with the sSFR.
This is what we see qualitatively in Fig.\,\ref{fig:flux_fraction_vs_MS} for substructures and measure as a weak correlation in Table\,\ref{tab:correlations flux with mass sfr}. 
For clumps, we find a slightly stronger correlation with the sSFR but, as already mentioned, the fraction of rest-frame NIR flux in clumps is much more strongly anti-correlated with the stellar mass of the host galaxy than its sSFR.
We note that Eq.\,\ref{eq:fraction of flux in NIR} should be rewritten as $\eta = F_{\rm s} / (F_{\rm o} + F_{\rm y} + F_{\rm s})$ for the most extended substructures since they can typically contribute to 5\% of the rest-frame NIR flux (sometimes up to 12\%; see the right panel of Fig.\,\ref{fig:clump_fraction}). For this type of substructures, the net effect is that an increase in $C_{\rm SFR}$ produces a shallower growth of $\eta$ than for clumps.

As noted previously, the fraction of rest-frame NIR flux in clumps decreases with increasing stellar mass at a given sSFR. If we interpret this in light of Eq.\,\ref{eq:fraction of flux in NIR}, this would mean that $C_{\rm SFR}$ would have to decrease with increasing stellar mass at a given SFR or sSFR, as is visible in Fig.\ref{fig:flux_fraction_vs_MS}. For instance, for galaxies in the MS at $z \approx 1.5$, this would require $C_{\rm SFR}$ to decrease by a factor of roughly 2.5 from $M_\star = 10^{9.75}\,\si{\Msun}$ to $10^{10.75}\,\si{\Msun}$.
Such a variation of $C_{\rm SFR}$ with the stellar mass of the host galaxy appears inconsistent with measurements of rest-frame UV clumps. For instance, in \citet{Guo2015}, they  find a contribution of UV clumps to the SFR between 5\% and 10\% without any significant difference between galaxies in the stellar mass range $10^9 < M_\star / \si{\Msun} < 10^{11.4}$.
Thus, it seems that a young stellar component cannot explain alone the fraction of NIR flux observed in clumps unless $C_{\rm SFR}$ is anti-correlated with the stellar mass of the galaxy host.

\subsection{Characteristics of multi-clump systems}
\label{sec:discussion/multi-clumps}

\begin{figure*}[htbp]
    \centering
    \includegraphics[width=\linewidth]{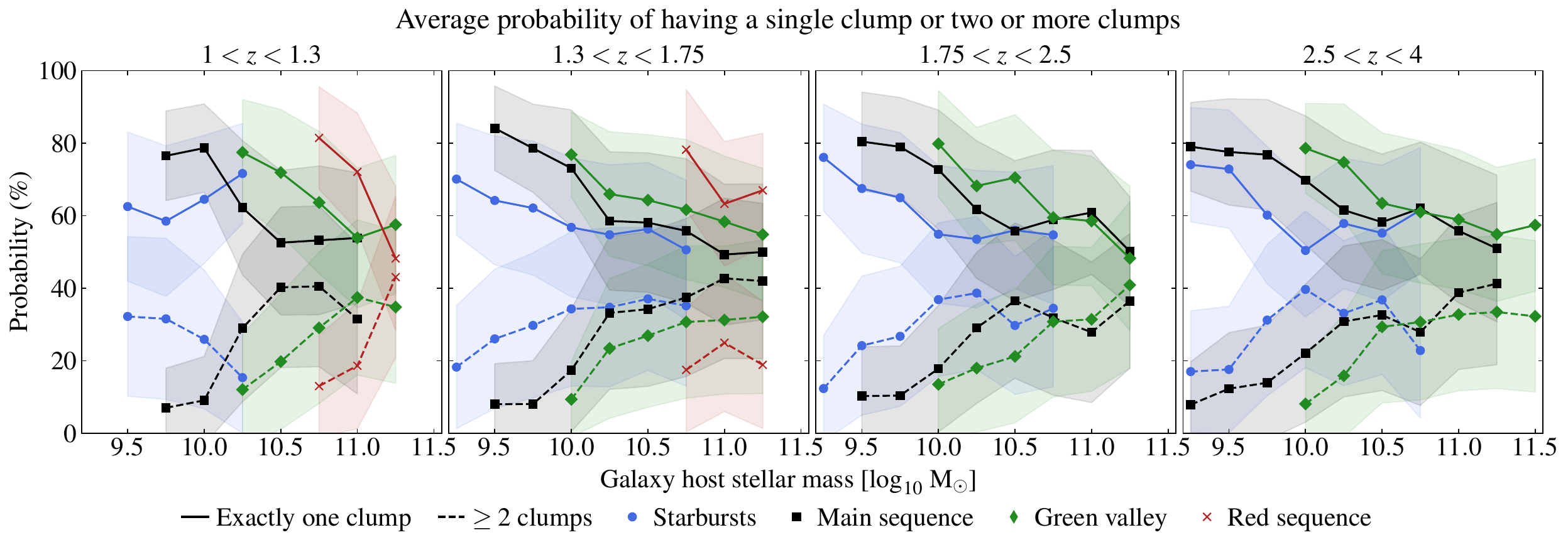}
    \caption{Average probability that a galaxy of a given stellar mass having at least one clump hosts either a single clump (continuous lines) or two or more clumps (dashed lines) for starbursts (blue circles), galaxies on the MS (black squares), in the green valley (green diamonds), and in the red sequence (red crosses). For each class, we show the standard deviation of the probability as a semi-transparent filled area with a similar color. 
    The probabilities are computed using Eqs.\,\ref{eq:appendix/probability more than n} and \ref{eq:appendix/probability exactly n}.}
    \label{fig:number_clumps}
\end{figure*}

We compare the physical properties of galaxies hosting a single clump with those hosting multiple clumps. The interested reader can find a similar discussion for rest-frame UV clumps in \citet{Ribeiro2017}. They find that about 25\% of their sample of star-forming galaxies is made of two-clump systems and favor major mergers as the main formation channel for such systems.
However, we note that our definitions of clumps are different (see Sects.\,\ref{sec:Results/overall/intrinsic} and \ref{sec:Results/size separation/intrinsic} for some of the key differences) which means that a direct comparison is not possible. Notably, they can detect the central bulge as a clump whereas we exclude it from our detections. The difficulty in comparing stems from the fact that the rest-frame UV is much more heavily affected by dust attenuation. It is therefore not clear how frequently the central bulge is detected as a clump in their analysis. This is important because if the bulge is detected as a clump in most cases, we should compare their two-clump systems to our single-clump systems. Otherwise, their two-clump systems are similar to ours and a direct comparison is possible.

To simplify the interpretation, we split the sample into starbursts, galaxies on the MS, in the green valley, and in the red sequence. We exclude galaxies within the region delimited by Eq.\,\ref{eq:delimiting line} since we do not find substructures in such galaxies.
We show in Fig.\,\ref{fig:number_clumps} the average probability (and its standard deviation) to find exactly one clump (continuous lines) or two or more clumps (dashed lines) for a given type of galaxy and a given stellar mass in four redshift bins spanning each roughly \SI{1}{\giga\year} of cosmic time. We consider galaxies for which we detect at least one clump to compute the average probability using Eqs.\ref{eq:appendix/probability more than n} and \ref{eq:appendix/probability exactly n}.
The probabilities of having one and two or more clumps do not sum to 100\% because galaxies also have a non-zero probability of hosting no clump. We note that if we select galaxies with at least one clump by using a cut on the probability $\Prob{N \geq 1}$ (see Appendix\,\ref{appendix:weighting scheme} for the notation) we recover the same trends.

Whatever the redshift, stellar mass, and type of the galaxy host, the population of clumpy galaxies is dominated by galaxies hosting a single clump. In the low-mass regime ($M_\star = 10^{9.5}\,\si{\Msun}$), the probability is higher in the MS (about 85\%) than in the starburst region (about 60\%). We do not see a strong variation with redshift, except at $2.5 < z < 4$ where the probabilities for galaxies on the MS and in the starburst region are both equal to approximately 80\%.
At higher stellar masses and for all galaxy types, the probability to find a single clump in a clumpy galaxy decreases, reaching 60\% at $z > 1.3$ and $M_\star \approx 10^{10.5}\,\si{\Msun}$ for starbursts and galaxies on the MS. For galaxies on the MS at $M_\star \geq \SI{e11}{\Msun}$, the probability to have a single clump reduces to nearly 50\%. 
Furthermore, we find that the probability that galaxies in the green valley have a single clump is systematically above that of galaxies on the MS. For instance, at $1 < z < 1.3$, galaxies on the green valley with $M_\star \approx 10^{10.5}\,\si{\Msun}$ have a probability of roughly 75\% while those on the MS have 55\%.
Similarly, massive galaxies in the red sequence ($M_\star > 10^{10.5}\,\si{\Msun}$) systematically have a higher probability to host a single clump than other types of galaxies.

A consequence of looking at the population of clumpy galaxies is that if the probability of hosting a single clump decreases with stellar mass, then the probability of hosting two or more clumps must evolve in the opposite way. This is what we find in Fig.\,\ref{fig:number_clumps}. Whatever the redshift, the probability to find a clumpy galaxy with multiple clumps is below 10\% for galaxies on the MS and in the green valley at low stellar masses ($M_\star < \SI{e10}{\Msun}$). In this mass range, starbursts are the only galaxies with a non-negligible probability ($\sim 30\%$) to have two or more clumps.
Whatever the type of galaxy, the probability to find two or more clumps strongly increases with the host's stellar mass, reaching values as high as 40\% at the high-mass end of the MS and the green valley. Even, galaxies on the red sequence at $z < 1.3$ have a probability close to 50\% to host two or more clumps.

We further check whether it is more probable for galaxies to host exactly two clumps.
For galaxies in the red sequence, we find that the majority of multi-clump systems host exactly two clumps. For starbursts and galaxies on the MS and in the green valley, we find that the probability to find exactly two clumps is close to the values presented in Fig.\,\ref{fig:number_clumps}, on average reduced by ten percentile points. This means that massive starbursts and galaxies on the MS have a probability to host exactly two clumps between 20\% and 30\% which is consistent with the fraction of galaxies with exactly two rest-frame UV clumps derived by \citet{Ribeiro2017}. 
Therefore, single-clump systems dominate the population of both star-forming and quiescent galaxies at all redshifts and stellar masses. Multiple-clump systems (including mostly two-clump ones) are nevertheless non-negligible. They are more commonly found in star-forming galaxies than in quiescent ones and they contribute more importantly to the clumpy population at higher stellar masses, in particular at the high-mass end of the MS and green valley.

\subsection{Interpreting the origin of rest-frame NIR substructures}
\label{sec:discussion/formation pathways}

Two major pathways regarding the origin of giant clumps in high-redshift galaxies are typically discussed. 
On the one hand, clumps could be produced by local disk fragmentation in turbulent and gas-rich disks \citep[e.g.][]{Elmegreen2006, Genzel2008, Genzel2011}. If clumps can survive over sufficiently long time-scales (typically \SI{1}{\giga\year} or above), their rest-frame NIR flux would decouple from the current SFR of their host galaxy. This would explain the lack of correlation between the fraction of rest-frame NIR flux measured in flux clumps and the galaxy host's SFR, as discussed in Sect.\,\ref{sec:Results/MS/flux}.
Furthermore, the fact that massive galaxies with high SFR values (${\rm SFR} \gtrsim \SI{10}{\Msun\per\year}$) are the most likely to host multiple clumps is also expected from disk fragmentation.
However, we note that certain simulations \citep[e.g.][]{Bournaud2014} suggest that the old stellar component of long-lived giant clumps is partially disrupted by tidal stripping while a young stellar component is continuously replenished by star formation. In such a case, we would expect to see come correlation with the SFR of the galaxy host.

On the other hand, the observed clumps could be the remnants of past galaxy minor \citep[e.g.][]{Guo2015} or major mergers \citep[e.g.][]{Ribeiro2017}. This would be consistent with clumpy galaxies hosting usually a single clump that would correspond to the surviving core of the secondary galaxy that has merged. However, depending on the typical merging time-scale, the frequency of mergers, and on the migration time-scale of the remaining core within the primary galaxy, we may expect to detect multiple cores as clumps. Therefore, the existence of multi-clump systems does not necessarily favor fragmentation over galaxy mergers as the main formation channel for clumps.
In addition, we expect galaxy mergers and fly-byes to produce tidal features that we may detect as extended substructures. If we believe that galaxy mergers produce a significant fraction of starbursts \citep[e.g.][]{Faisst2025}, then we would expect extended substructures to dominate in that population. This appears to be consistent with Fig.\,\ref{fig:flux_fraction_vs_MS} in which we see that the fraction of rest-frame NIR flux in substructures peaks in the starburst region, but not for clumps, thus showing that extended substructures dominate the flux budget in that population of galaxies.
Nevertheless, we note that extended substructures do not have to be tidal tails but, instead, could be bars and spiral arms produced by density waves or stochastic clumpy star formation, as suggested by \citet{Kalita2025_spiral}. 

\begin{figure}
    \centering
    \includegraphics[width=\linewidth]{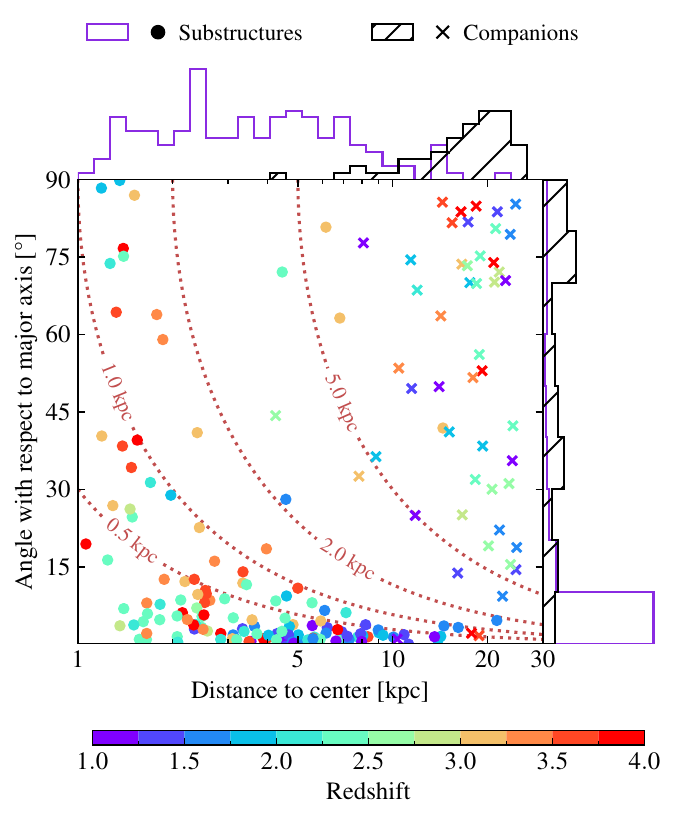}
    \caption{Position of substructures (circles) with respect to the galaxy host color-coded with redshift. Substructures are detected with the intrinsic detection method in edge-on galaxies with bulges contributing less than 50\% to the rest-frame NIR flux. We show on the x-axis the distance to the center of the host galaxy and, on the y-axis, the angle between the barycenter of the substructure and the major axis of the host galaxy. We over-plot with crosses the distribution of galaxy companions with properties comparable to clumps. Companions are selected at a distance less than \SI{3}{\arcsec} from the main galaxy, with a total flux in the rest-frame NIR contributing less than 15\%, and within $1\sigma$ of the redshift of the main galaxy.
    We overlay with pink dotted lines iso-contours of the vertical distance to the galaxy plane.}
    \label{fig:position of clumps}
\end{figure}

Disentangling between clumps formed in-situ (i.e., fragmentation) and ex-situ (i.e., mergers) is not trivial. One way to determine which origin is more likely is by looking at the spatial distribution of clumps. 
Indeed, clumps produced via disk fragmentation should be formed and then migrate inside the stellar disk. On the other hand, if the rest-frame NIR clumps trace galaxy mergers, we expect to find a significant fraction of them outside of the plane of the disk component of the galaxy.
We use clumpy edge-on galaxies to study the spatial distribution of clumps. We select the sub-sample by requiring that the galaxies have a stellar disk axis ratio $b/a < 0.15$, with $b$ and $a$ the minor and major axes, respectively, and a bulge-to-total rest-frame NIR flux ratio ${\rm B/D} < 0.5$. The latter criterion is necessary to ensure selecting galaxies that host a disk component that is bright enough, otherwise the stellar disk axis ratio would become loosely constrained. This yields a sub-sample of 85 clumpy edge-on galaxies, more than 80\% of which are starbursts or on the MS.
Finding substructures in edge-on galaxies is biased because we detect them within the extent of the segmentation maps produced by \SEpp{}. Since the segmentation maps tend to follow the shape of the galaxies, this disfavors detecting substructures along the minor axis if they are located sufficiently far from the disk plane. In turn, this can give the impression that substructures are located preferentially within the plane of the disk. Luckily, if a substructure located outside of the disk plane is not associated to the segmentation map of the galaxy, that means it was detected as a separate galaxy by \SEpp{}. Therefore, we can use the \COSMOScat{} catalog from \citet{Shuntov2025} to search for nearby companions to edge-on galaxies and include them in our analysis. To get companions comparable to substructures, we require that
\begin{enumerate*}
    \item they are located at most \SI{3}{\arcsec} away from the center of the main galaxy\footnote{This is the extent of the \JWST{} stamps in which the substructure detection is carried out.},
    \item their photometric redshift is within $1\sigma$ of the redshift of the main galaxy, and
    \item their contribution to the rest-frame NIR flux is at most 15\%.
\end{enumerate*}
The latter criterion is used to remove galaxy companions that are too bright to be consistent with the fluxes measured in Sect.\,\ref{sec:Results/MS/flux}. With this criterion, the median rest-frame NIR flux fraction in companions is roughly 5\%, with the 16th and 84th percentiles equal to 1.75\% and 10\%, respectively.

We show in Fig.\,\ref{fig:position of clumps} the spatial distribution of substructures detected with the intrinsic detection method (circles) and galaxy companions (crosses) color-coded by the redshift of the host galaxy. Along the x-axis, we represent the distance to the center of the galaxy and, along the y-axis, the angle between the barycenter of the clump (or companion) and the major axis of the host, modulo \SI{90}{\degree}. We do not separate between clumps, moderately large, and extended substructures because they are distributed similarly. Since a given physical distance can carry different meanings depending on the size of the galaxies, we provide in Fig.\,\ref{fig:position of clumps normalised} a similar figure but with the distance to the center normalized by the scale length of the stellar disk\footnote{The disk scale length is defined as $R_{\rm d} = R_{\rm eff, d} / b_1$, with $R_{\rm eff}$ the effective radius measured by \SEpp{} on \JWST{}/NIRCam images and $b_1 \approx 1.6783$ \citep{Graham_2005}.}. 

We find that half of substructures are located closer than \SI{3.5}{\degree} from the disk plane, and 75\% are closer than \SI{10}{\degree}. Furthermore, half of them are located less than \SI{3.5}{\kilo\parsec} away from the center of their host galaxy and 75\% are within approximately \SI{6}{\kilo\parsec}\footnote{We remind that we do not detect substructures below \SI{1}{\kilo\parsec}.}. If we measure the vertical distance of substructures with respect to the plane of the stellar disk, we find that half of substructures are closer than \SI{250}{\parsec} and 75\% are below \SI{650}{\parsec}. This latter distance is close to the scale-height measurements of disk galaxies at $z \approx 1$, approximately equal to \SI{500}{\parsec} \citep[e.g.][]{Usachev2024}. In other terms, the majority of substructures detected in edge-on galaxies appear to be located within the plane of the stellar disk. 
We note that the distributions of substructures above and below $z = 3$ are different. At high-redshift, substructures appear closer to the center of the galaxy and about 25\% of them are distributed with an angle larger than \SI{30}{\degree} (8\% at $z < 3$) and with a vertical distance ranging from \SI{500}{\parsec} to \SI{2}{\kilo\parsec}. 
Interestingly, when normalizing the distance by the scale length of the stellar disk, the difference between low and high redshift becomes less clear, with a median distance of $1.5 R_{\rm d}$ and the 16th and 84th percentiles equal to $0.6 R_{\rm d}$ and $2.6 R_{\rm d}$, respectively. There is however a population of substructures in galaxies at $z > 2$ that are found both at small ($< 0.5 R_{\rm d}$) and large distances ($5 R_{\rm d}$) and that we do not find at lower redshift.
On the other hand, galaxy companions are distributed relatively evenly in terms of angle which confirms the hypothesis that if galaxy mergers produce clumps, they should be distributed in all directions. Furthermore, companions are located at large distances, typically beyond \SI{5}{\kilo\parsec} from the center, at $R > 5 R_{\rm d}$, and at large vertical distances, with 80\% of them beyond \SI{5}{\kilo\parsec}.

There is a region in Fig.\,\ref{fig:position of clumps} between approximately \SI{2}{\kilo\parsec} and \SI{5}{\kilo\parsec} above the plane of the stellar disk where very few substructures and companions are found. This could be a dynamical effect due to the nature of mergers. If the secondary galaxies have eccentric orbits, they are expected to spend more time close to their apocenter where they would be detected as companions located at large distances. On the other hand, during their last phase of merger, if they merge in an orbit that is not coplanar with the stellar disk of the main galaxy, they would be identified as substructures close to the center and with a large angle.
However, we note that it is possible that, after a few orbits, the dynamical interactions between the two galaxies bring the galaxy companion into the stellar disk plane, in which case the merger remnants would be identified as substructures. Since roughly 11\% of edge-on galaxies have at least one companion in a shell between \SI{10}{\kilo\parsec} and \SI{30}{\kilo\parsec}, we can estimate the fraction of companions that would end-up detected as substructures between \SI{1}{\kilo\parsec} and \SI{10}{\kilo\parsec} if they all move into the stellar disk. We find that about 1.5\% of edge-on galaxies would host a merger companion detected as a clump. In comparison, when combining all redshift ranges, we find that 21\% of edge-on galaxies host at least one substructure.
Therefore, galaxy mergers do not seem to be the driving mechanism behind the formation of substructures in galaxies. On the other hand, local disk fragmentation is a much more likely scenario. This conclusion appears consistent with the one reached by \citet{Huertas-Company2020} for clumps detected in the rest-frame UV and optical.

\section{Conclusion}

We have studied substructures at a rest-frame wavelength of \SI{1}{\micro\meter} in galaxies at $1 < z < 4$ with $M_\star > \SI{e9}{\Msun}$ in the \CWeb{} survey by identifing contiguous substructures in the residual images from the F277W and F444W \JWST{}/NIRCam bands, following the removal of an axially symmetric bulge-disk model. We have used two distinct methods:
\begin{enumerate*}
    \item an optimal approach whereby every substructure above a given flux level is detected and
    \item an intrinsic approach whereby redshift dependent surface and flux detection thresholds are applied to correct for cosmologically-induced biases.
\end{enumerate*}
 Each approach allows us to find small substructures (i.e. clumps), moderately large ones, and extended ones. Regarding the optimal detection method, we find the following results.

\begin{enumerate}
    \item Rest-frame NIR substructures are ubiquitous in galaxies at $z = 1$ and are present in half of them at $z = 4$. And
    \item clumps are the most common type of substructures at all redshifts, though extended substructures are found in approximately 20\% of galaxies at $z = 1$ (a few percents at $z = 4$).
\end{enumerate}

The caveat with the optimal detection approach is that we do not detect substructures that are comparable in size and luminosity at different redshifts, thus biasing our results. 
Using the intrinsic approach, we correct for the redshift dependence of the size and flux of substructures, and we find the following results.

\begin{enumerate}
    \item The galaxy population, as a whole, has become less clumpy over cosmic time, with the fraction of galaxies with substructures going from about 40\% at $z = 4$ to 10\% at $z = 1$.
    \item Clumps are the most common type of substructures at all redshifts, even though moderately large substructures contribute to about half at $z = 1$. And
    \item the fraction of rest-frame NIR flux found in substructures is small (1\% for clumps, 3\% for moderately large substructures, 5\% for extended substructures) compared to the fractions of rest-frame UV flux found in clumps that are quoted in the literature (typically more than 30\%). Furthermore, we do not measure an evolution with redshift.
\end{enumerate}

Since our sample comprises both star-forming and quiescent galaxies, we have checked in which population substructures are more likely to be located by looking at the prevalence of substructures and their fraction of flux in the rest-frame NIR across the $M_\star - {\rm SFR}$ plane. We find the following.

\begin{enumerate}
    \item Rest-frame NIR substructures are absent in galaxies with low stellar mass and SFR ($M_\star \lesssim 10^{10.5}\,\si{\Msun}$ and ${\rm SFR} \lesssim \SI{10}{\Msun\per\year}$).
    \item Substructures are ubiquitous in massive galaxies ($M_\star \gtrsim \SI{e10}{\Msun}$) on the MS, in the starburst region and at the high-SFR end of the green valley. Similarly, approximately 70\% of that population of galaxies hosts at least one clump.
    \item The brightest substructures (approximately 4\% of the rest-frame NIR flux) are found in galaxies with the highest sSFR values and the faintest ones are found at the high-mass end the red sequence. However, the brightest clumps are found in low-mass galaxies ($M_\star \approx \SI{e9}{\Msun}$) and the faintest clumps are found in the most massive galaxies on the MS, starburst region, and green valley. And
    \item at a given location in the $M_\star - {\rm SFR}$ plane, we do not measure any significant redshift evolution of the fraction of galaxies with substructures or clumps. One exception is at the low-SFR end of the green valley, near the transition to the red sequence, where the fraction galaxies with substructures has dropped by 20 percentile points from $z = 4$ to $z = 1$.
\end{enumerate}
We conclude that, whatever the redshift, the clumpiness of the galaxy population in the rest-frame NIR is defined by the stellar mass and SFR of the host galaxies. Thus, as galaxies evolve along the MS or reach the starburst region, they are more likely to develop substructures, including clumps. On the other hand, when galaxies quench their star-formation, they appear less likely to retain substructures.
We interpret the reduction of the fraction of clumpy galaxies with cosmic time as being induced by the evolution of the entire galaxy population within the $M_\star - {\rm SFR}$ plane towards lower SFR values and by the build-up of the low-mass green valley and red sequence after the peak of star formation at $z \approx 2$.

Overall, we find that clumpy galaxies are more likely to host a single clump. This is especially true at low stellar masses ($M_\star < \SI{e10}{\Msun}$) where clumpy galaxies on the MS have a 80\% probability to have one clump (between 60\% and 80\% for starbursts). Incidentally, this means that low-mass starbursts are more likely to host multiple clumps (probability of 30\%) than galaxies on the MS (a few percents).
Moreover, we find that massive clumpy galaxies are more likely to host multiple clumps, up to 40\% at $M_\star \approx \SI{e11}{\Msun}$ for those on the MS and in the green valley. 

By looking at the spatial distribution of rest-frame NIR substructures in edge-on galaxies, we find that they are mostly located close to the stellar disk plane. Such a spatial distribution is expected from local disk fragmentation but could also be the result of galaxy mergers if the merger remnants end up preferentially in the disk. We use the detections of nearby galaxy companions with properties comparable to the detected substructures to evaluate the fraction of substructures that could be the result of past galaxy mergers. Overall, we find that galaxy mergers are insufficient to explain the bulk of substructures and that local disk fragmentation is likely the most plausible scenario behind the formation of such rest-frame NIR substructures, and in particular clumps.

We note that this conclusion only applies to disk galaxies. Therefore, clumps found in other populations could still be produced ex-situ. In particular, this could be the case for clumpy galaxies at the high-mass end of the red sequence that tend to host a single clump in the majority of cases. More detailed studies of substructures in different populations of galaxies are required to understand their origin. In particular, one could
\begin{enumerate*}
    \item evaluate the physical properties of substructures using SED modeling (e.g. stellar mass, SFR, or age) and correlate them to the characteristics of their host galaxy \citep[for an attempt and discussions therein, see][]{claeyssens_star_2023, Claeyssens2024, Kalita2024, Kalita2025},
    \item study the presence and properties of substructures as a function of the SFH of their galaxy host. Typically, this could be used to put constraints on the survivability timescale of clumps. And
    \item one could compare our detections of substructures with mocks \CWeb{} observations produced from simulations to better constrain the physics leading to the formation of realistic rest-frame NIR clumps and substructures.
\end{enumerate*}

\begin{acknowledgements}
This work made use of \texttt{python} and the following packages: \texttt{sqlite3} \citep{sqlite2020hipp}, \texttt{NumPy} \citep{numpy}, \texttt{Matplotlib} \citep{matplotlib}, \texttt{tqdm} \citep{tqdm}, \texttt{joblib}\footnote{\url{https://joblib.readthedocs.io/en/stable/}}, \texttt{scikit-image} \citep{scikit-image}, \texttt{scipy} \citep{2020SciPy-NMeth}, \texttt{pingouin} \citep{Vallat2018}, \texttt{Astropy} \citep{astropy:2022}, and its affiliated package \texttt{Regions} \citep{astropy_regions}.
This research is also partly supported by the Centre National d’Etudes Spatiales (CNES).
We acknowledge the funding of the French Agence Nationale de la Recherche for the project
iMAGE (grant ANR-22-CE31-0007).
We warmly acknowledge the contributions of the entire COSMOS collaboration consisting of more than 100 scientists. The HST-COSMOS program was supported through NASA grant HST-GO-09822. 
More information on the COSMOS survey is available at \url{https://cosmos.astro.caltech.edu}. 
This work was made possible by utilizing the CANDIDE cluster at the Institut d’Astrophysique de Paris, which was funded through grants from the PNCG, CNES, DIM-ACAV, and the Cosmic Dawn Center and maintained by S. Rouberol.
\end{acknowledgements}

%
%

\bibliographystyle{aa}
\bibliography{bib}

\appendix

\onecolumn
\section{Complementary figures}
\label{appendix:complementary figures}

\begin{figure}[htp]
    \begin{minipage}[c]{0.45\linewidth}
    \centering
    \includegraphics[width=\linewidth]{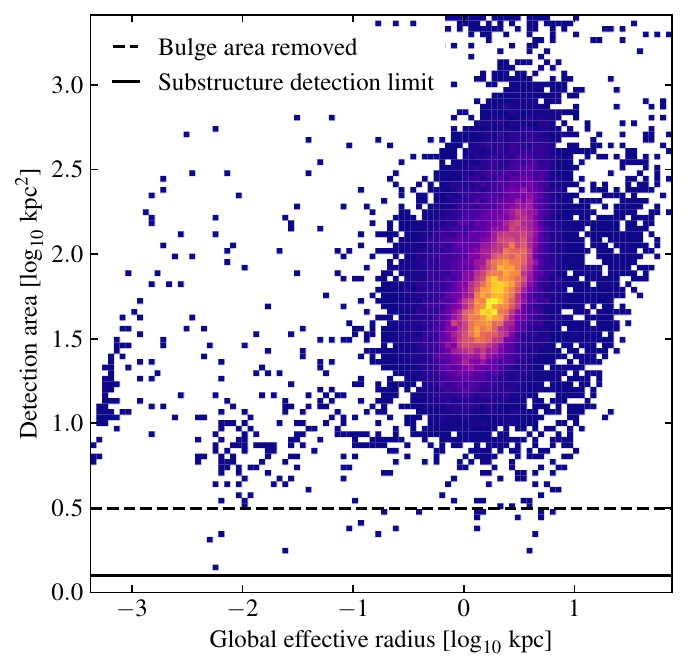}
    \caption{Density plot (low density in blue, high density in yellow) of the original area available for substructure detection as a function of the global effective radius of galaxies (see Eq.\,6 of \citealt{Mercier2022} for a definition). We represent the area associated with the removed bulge as a dashed line, while the continuous line indicates the threshold used for the intrinsic substructure detection.}
    \label{fig:detection area vs radius}
    \end{minipage}
    \hfill
    \begin{minipage}[c]{0.45\linewidth}
    \centering
    \includegraphics[width=\linewidth]{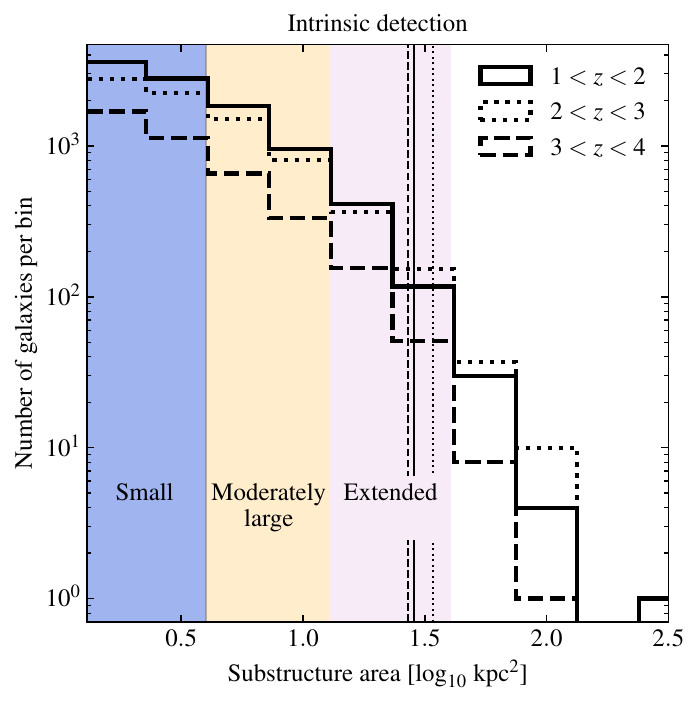}
    \caption{Distribution of substructures detected with the intrinsic method (see Sect.\,\ref{sec:detection/intrisic detection}) as a function of their area, divided into three redshift bins: $1 < z < 2$ (continuous line), $2 \leq z < 3$ (dotted line), and $3 \leq z < 4$ (dashed line). Vertical lines represent the 99\% quantile for each distribution. Background colors indicate the area bins used in the analysis: clumps (blue, $1.3 < S / \si{\kilo\parsec\squared} < 4$), moderately large substructures (orange, $4 < S / \si{\kilo\parsec\squared} < 13$), and extended ones (pink, $13 < S / \si{\kilo\parsec\squared} < 40$).}
    \label{fig:histogram area of substructures}
    \end{minipage}
    
    \begin{minipage}[c]{\linewidth}
    \centering
    \vspace{10pt}
    \includegraphics[width=\linewidth]{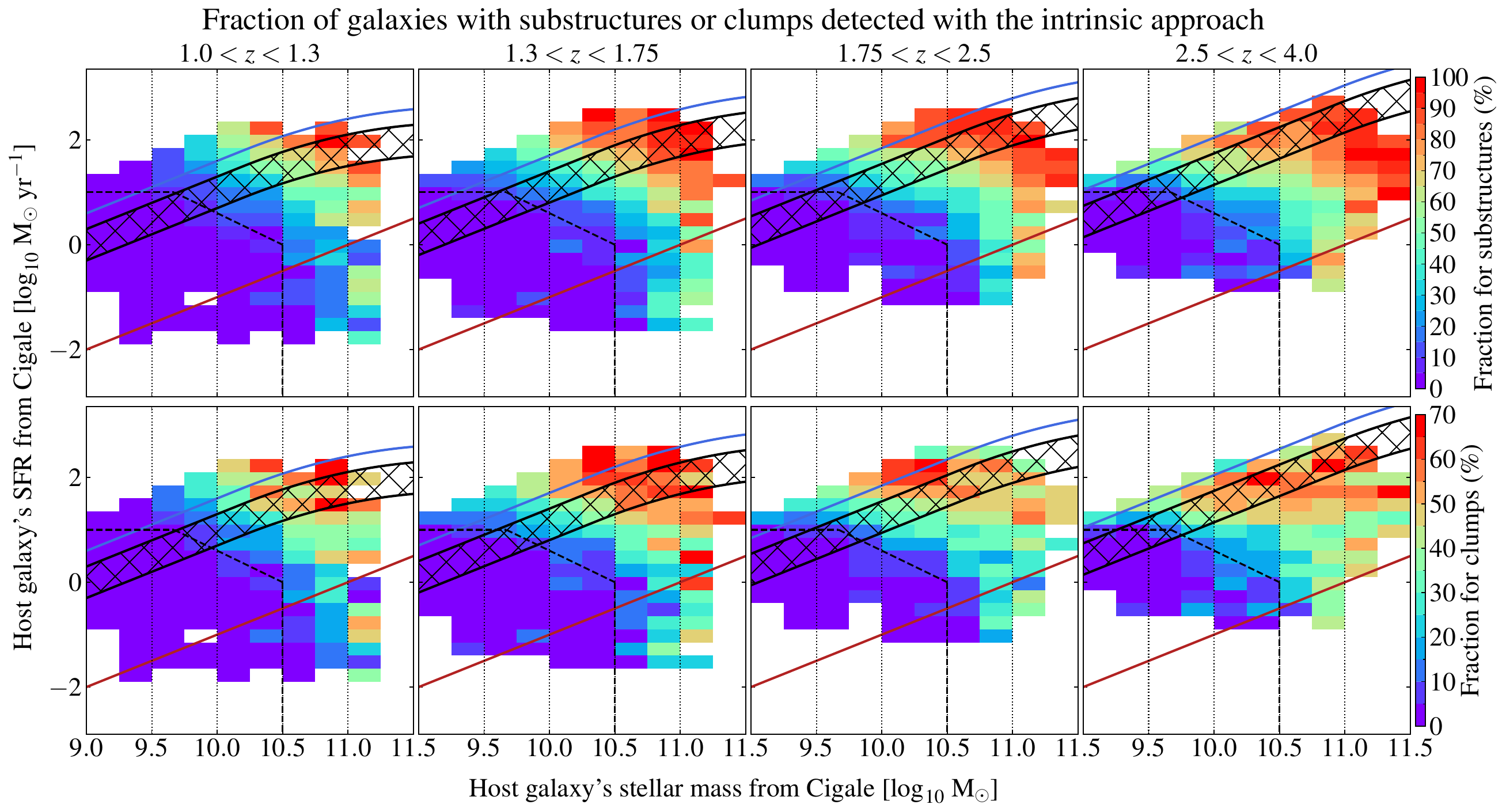}
    \caption{Same as Fig.\,\ref{fig:fraction_vs_MS} but using the stellar mass and SFR from \Cigale{}. We do not show bins with fewer than ten galaxies. The presence or absence of certain bins with respect to Fig.\,\ref{fig:fraction_vs_MS} is driven by the differences in stellar mass and SFR between \LEPHARE{} and \Cigale{}. All lines are defined similarly to Fig.\,\ref{fig:fraction_vs_MS}. This shows that our results are not impacted by the SED fitting tool and assumptions used in our analysis.}
    \label{fig:substructures vs MS with Cigale}
    \end{minipage}
\end{figure}

\begin{figure*}
    \centering
    \includegraphics[width=\linewidth]{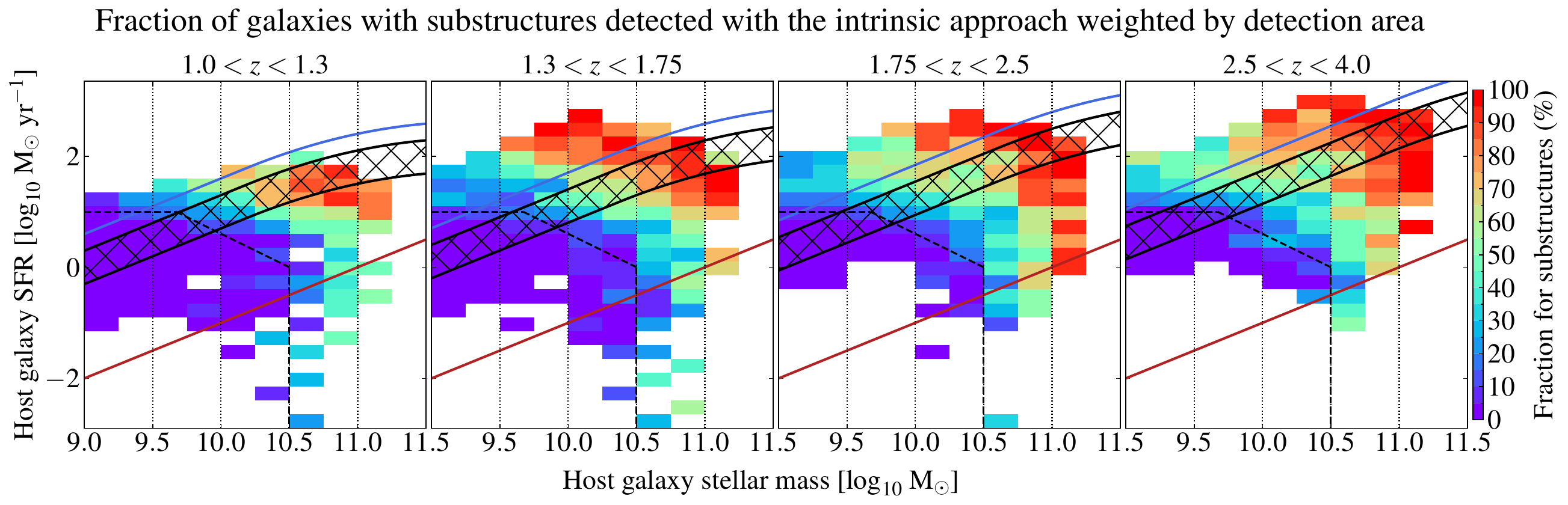}
    \caption{Same as the first row of Fig.\,\ref{fig:fraction_vs_MS} but weighting each galaxy by its detection area when computing, in each bin, the average fraction of galaxies with at least one substructure detected with the intrinsic approach. All lines are defined similarly to Fig.\,\ref{fig:fraction_vs_MS}. This shows that our results are not impacted by the detection area of substructures. The same applies to clumps.}
    \label{fig:substructures vs MS with area weigths}
\end{figure*}

\begin{figure*}
    \centering
    \includegraphics[width=0.95\linewidth]{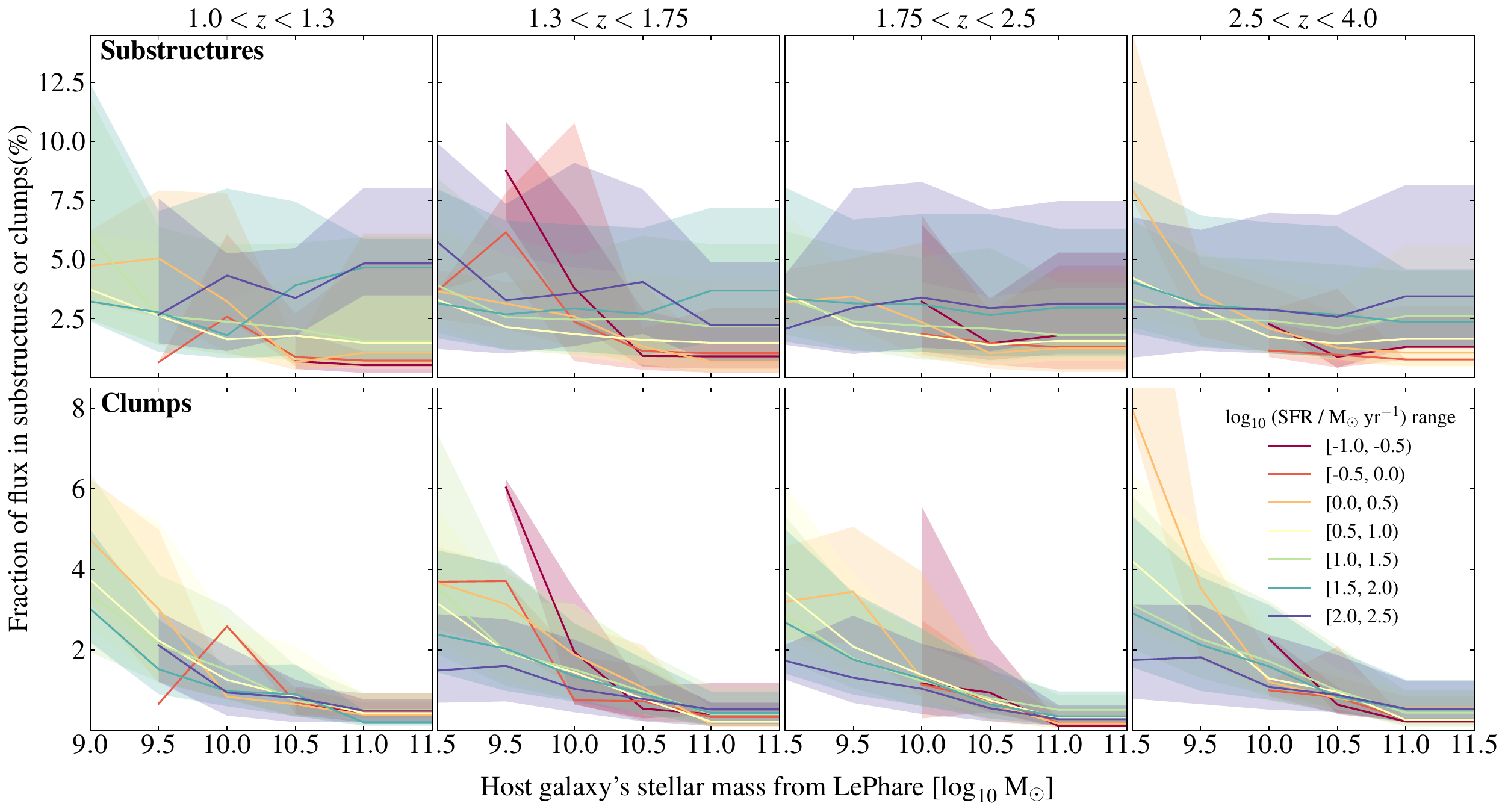}
    \caption{Fraction of rest-frame NIR flux in substructures (top row) and clumps (bottom row) as a function of the host galaxy's stellar mass in four redshift bins, each bin spanning \SI{1}{\giga\year} of cosmic history. Galaxies are split in different bins of SFR, as illustrated with the various colored lines. The shaded areas correspond to the 16th and 84th percentiles for each SFR bin. See Fig.\,\ref{fig:flux_fraction_vs_MS} for an alternative representation.}
    \label{fig:flux fraction vs mass with SFR as color}
\end{figure*}

\twocolumn
\begin{figure}
    \centering
    \includegraphics[width=\linewidth]{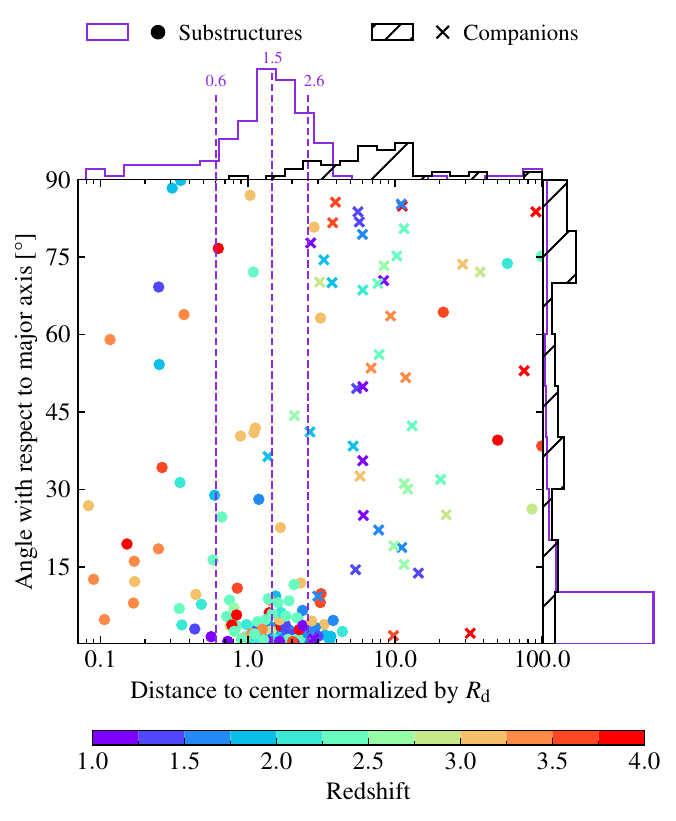}
    \caption{Figure similar to Fig.\,\ref{fig:position of clumps} but with the distance to the center of the host galaxy normalized by the disk scale length ($R_d$) derived with \SEpp{} on the four \JWST{}/NIRCam bands.}
    \label{fig:position of clumps normalised}
\end{figure}

\section{Details on substructure detection}
\label{appendix: detection}

To determine whether $2\sigma$ is sufficient for the optimal detection (similarly to what was done in previous works; e.g., $5\sigma$ in \citealt{Kalita2024a} or $4.5\sigma$ in \citealt{Kalita2024}), we have carried out a purity test. 
For each galaxy, we estimated iteratively the number $n$ of $\sigma$ that is required to not detect any structure in the background pixels\footnote{Performed with a dichotomy between $n = 0$ and $n = 10$ with steps of 0.1.}. We find a mean value of $n = 1.4$, a median value of $n = 1.5$, with the first quartile at $n = 1.3$, the third quartile at $n = 1.6$, and the 99\% quantile at $n = 2.1$. We therefore settled on a flux threshold of $2\sigma$. 
The primary reason for the difference in thresholds between this paper and other ones \citep[e.g.][]{Kalita2024a, Kalita2024} is likely that they exclude substructures, which results in higher thresholds to isolate bright clumps located on top of fainter but more extended substructures.

\begin{figure}[htbp]
    \centering
    \includegraphics[width=0.95\linewidth]{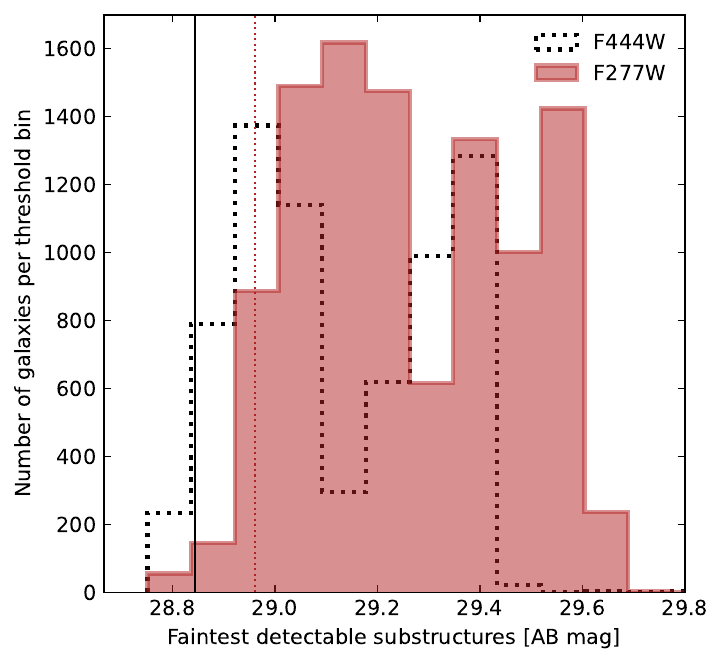}
    \caption{Histogram of the faintest detectable substructure in each galaxy. We split between galaxies for which the detection was carried out in the \JWST{}/NIRCam F277W band ($1 < z < 2$; filled in red) and in the F444W band ($2 \leq z < 4$; black dotted line). Vertical lines represent the 5\% quartiles for the two bands, that is the values such that 95\% of the galaxies have a lower detection threshold. We note that one must add \SI{3.25}{\mag} (or equivalently divide the corresponding flux threshold by 20) to convert these values into the detection threshold applied on each pixel.}
    \label{fig:detection threshold}
\end{figure}

We show in Fig.\,\ref{fig:detection threshold} the histogram of the faintest substructures that can be detected for the entire sample. We illustrate galaxies at $1 < z < 2$ in red (detection in F277W) and those at $2 \leq z < 4$ with a black dotted line (detection in F444W). As expected, the detections in F277W (the deepest band, see \citealt{CaseyCWebSurvey}) are slightly deeper than in F444W band with the 5\% quantile (i.e., 95\% of the galaxies have a lower threshold) at \SI{28.85}{\mag} in F444W and at \SI{28.95}{\mag} in F277W.

\begin{figure}[htbp]
    \centering
    \includegraphics[width=0.95\linewidth]{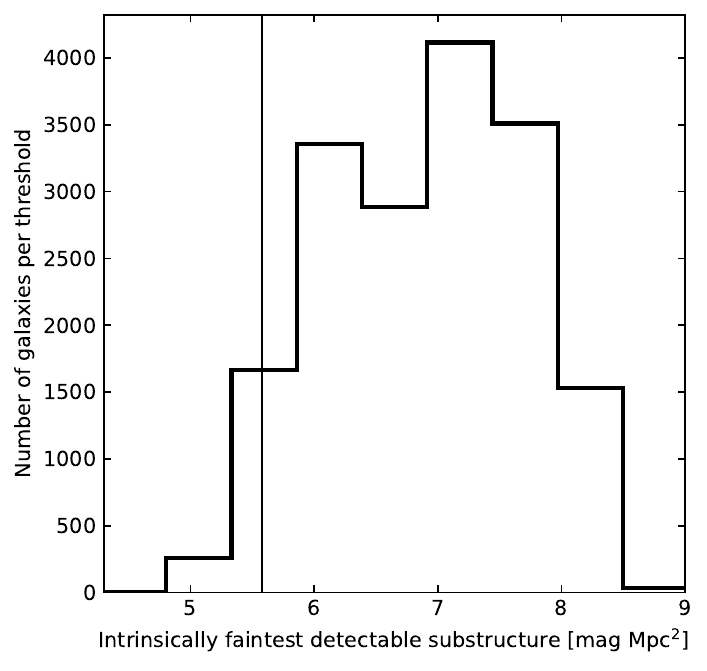}
    \caption{Histogram of the intrinsically faintest detectable substructure in each galaxy. The intrinsic flux threshold is computed for a substructure of 20 pixels with a $2\sigma$ detection per pixel, with $\sigma$ measured from the local background around each galaxy (see Sect.\,\ref{sec:detection/maximal detection}),  multiplied by $4\pi D_{\rm L}^2 / (1+z)$ to take out the redshift dependence of the flux density. The vertical line marks the 5\% quartile of the distribution, that is the value such that 95\% of the galaxies have an intrinsic flux detection threshold that is higher than this value.}
    \label{fig:intrinsic detection threshold}
\end{figure}

\begin{figure}[htbp]
    \centering
    \includegraphics[width=0.95\linewidth]{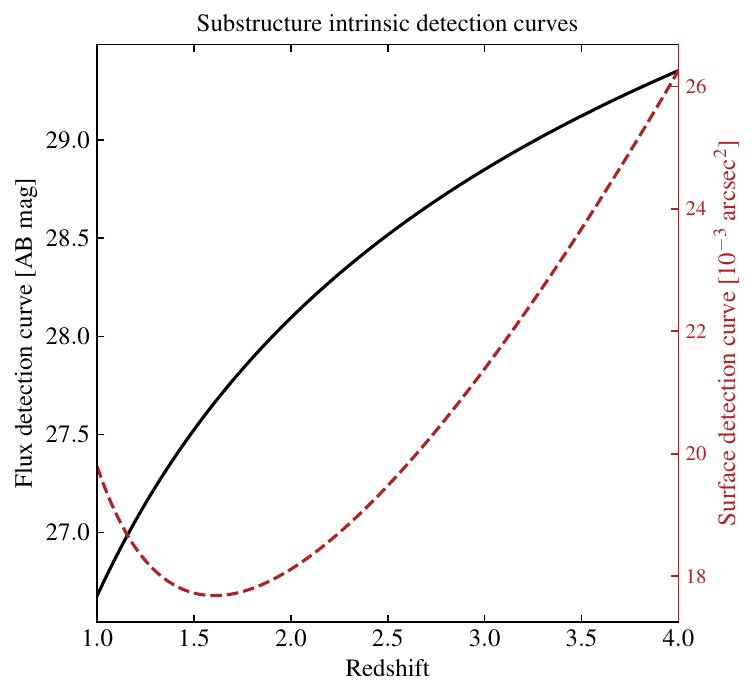}
    \caption{Flux and surface detection curves derived in Sect.\,\ref{sec:detection/intrisic detection} to select substructures with flux and area values that are above thresholds independent of redshift. The black continuous line represents the flux detection curve and the dashed red line represents the surface detection curve. Similarly to Fig.\,\ref{fig:detection threshold}, one needs to add \SI{3.25}{\mag} to this curve to obtain the flux detection curve applied to each pixel individually during the substructure detection process.}
    \label{fig:detection curves}
\end{figure}

\begin{figure}[htbp]
    \centering
    \includegraphics[width=\linewidth]{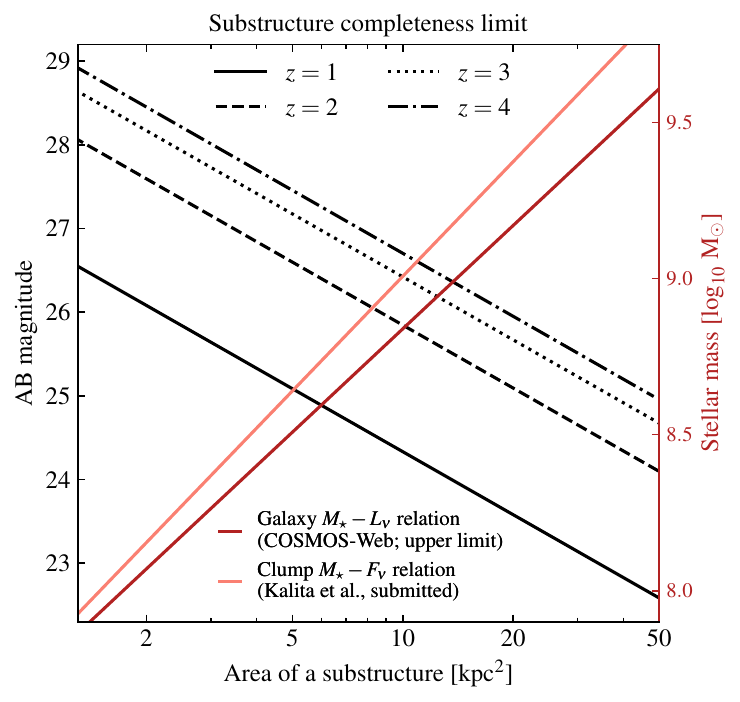}
    \caption{Theoretical completeness limit in terms of magnitude (left vertical axis and black lines) and stellar mass (right vertical axis and red lines) for the intrinsic detection method as a function of the area of the substructures and of their host galaxy's redshift. The stellar mass completeness is evaluated at $z = 1.5$ and is expected to increase by \SI{0.2}{\dex} at $z = 4$. We estimate the stellar mass completeness, first, using the best-fit relation between specific luminosity at a rest-frame wavelength of \SI{1}{\micro\meter} and stellar mass for the entire \CWeb{} sample (dark red line) and, second, using the best-fit relation from \citet{Kalita2025} for clumps in star-forming galaxies at $z \approx 1.5$ detected in rest-frame optical (salmon-colored line).
    }
    \label{fig:completeness}
\end{figure}

\clearpage
\section{Weighted fraction of galaxies with substructures}
\label{appendix:weighting scheme}

\subsection{Probability of substructure association with a galaxy}

In some cases, substructures detected in the residual maps fall in-between two or more galaxies, making it difficult to associate them to one or the other. Instead of discarding such cases, we implement a weighting scheme when computing the fraction of galaxies with one or more substructures, as follows.
For each galaxy, we loop through its substructures and search for nearby galaxies within a circular radius of \SI{6}{\arcsec}. This corresponds to the size of the cutouts used to detect substructures and we find that it is sufficient to detect nearby companions to the main galaxy. Let us assume that $N$ galaxies $G_i$ are detected for substructure $s$. Each galaxy will be separated by an angle $\theta_i$ from the substructure. To determine the angular separation, we use the center of the galaxies as well as the median right ascension and declination of the substructures as derived from their segmentation map.
For each galaxy $G_i$, we must associate a probability $\Prob{s \in G_i}$ that the substructure $s$ belongs to it. We define this probability as follows:

\begin{align}
    \label{eq:appendix/definition_prob1}
    \forall i, j,\ \ \theta_i\,\Prob{s \in G_i} &= \theta_j\,\Prob{s \in G_j}, \\
    \sum_{i = 1}^N \Prob{s \in G_i} &= 1
    \label{eq:appendix/definition_prob2}
\end{align}

Simply put, Eq.\,\ref{eq:appendix/definition_prob1} states that a substructure twice closer to a galaxy should have its probability doubled and Eq.\,\ref{eq:appendix/definition_prob2} that a substructure must belong to one of the galaxies. Assume we want to determine the probability $\Prob{s \in G_k}$. We can start from Eq.\,\ref{eq:appendix/definition_prob2} and use Eq.\,\ref{eq:appendix/definition_prob1} to isolate $\Prob{s \in G_k}$, which leads to

\begin{equation}
    \Prob{s \in G_k} = \left [ \sum_{i = 1}^N \theta_k / \theta_i \right ]^{-1},
\end{equation}
at which point Eq.\,\ref{eq:appendix/definition_prob1} can be used to quickly calculate all other probabilities.

\subsection{Probability of a galaxy having at least one substructure}

What really interests us is the probability that a galaxy possesses at least one substructure\footnote{We note that this also applies to sub-classes of substructure such as clumps.}. Let us call $N_s$ the number of substructures that a galaxy holds. The probability we are interested in therefore writes $\Prob{N_s \geq 1} = 1 - \Prob{N_s = 0}$.
Let us further assume that a galaxy $G$ has $N$ detected substructures $s_i$ each with a probability $\Prob{s_i \in G}$. $\Prob{N_s = 0}$ means that the galaxy has no substructures, that is that all $s_i$ do not belong to $G$:

\begin{equation}
    \Prob{N_s = 0} = \Prob{\bigwedge_{i=1}^N s_i \notin G },
    \label{eq:appendix/probability of no clump}
\end{equation}
where $\wedge$ is the logical conjunction. The fact that one substructure belongs to $G$ should not affect any other substructure. In other terms, we can assume that all variables $s_i$ are independent from one another which simplifies Eq.\,\ref{eq:appendix/probability of no clump} into a product of $\Prob{s_i \notin G} = 1 - \Prob{s_i \in G}$ and leads to the final formula:

\begin{equation}
    \Prob{N_s \geq 1} = 1 - \prod_{i=1}^N \left [ 1 - \Prob{ s_i \in G } \right ].
    \label{eq:appendix/probability of at least one clump}
\end{equation}

\subsection{Probability of a galaxy having at least $n$ substructures}

Another related probability that is interesting to compute is $\Prob{N_s \geq n}$. This writes:

\begin{equation}
    \forall n > 0, \ \ \Prob{N_s \geq n} = 1 - \sum_{k=0}^{n-1} \Prob{N_s = k}.
    \label{eq:appendix/probability more than n}
\end{equation}

An alternative is to write it down in terms of $\Prob{N_s \geq n-1} - \Prob{N_s = n-1}$ and to solve recursively, but that requires calculating the same $\Prob{N_s = k}$ terms. The condition $N_s = k$ requires computing the union of the combinations of $k$ substructures that belong to $G$, and of the $N - k$ ones that do not. Let us call $\mathcal{S} = \lbrace s_i \rbrace$ the set of $N$ substructures and $\left ( \mathcal{S}, k \right )$ the set of $k$-combinations of $\mathcal{S}$. Then we have

\begin{equation}
    \Prob{N_s = k} = \Prob{\bigvee_{\mathcal{L} \in (\mathcal{S}, k)} \left [ \bigwedge_{s \in \mathcal{L}} s \in G \bigwedge_{s \in \mathcal{S} \backslash \mathcal{L}} s \notin G \right ]},
\end{equation}
where $\vee$ is the logical disjunction and $\mathcal{S} \backslash \mathcal{L}$ is the set difference between $\mathcal{S}$ and $\mathcal{L}$. Assuming the independence of all $s_i$, this simplifies as

\begin{equation}
    \Prob{N_s = k} = \sum_{\mathcal{L} \in (\mathcal{S}, k)} \left [ \prod_{s \in \mathcal{L}} \Prob{s \in G} \prod_{s \in \mathcal{S} \backslash \mathcal{L}} \left ( 1 - \Prob{s \in G} \right ) \right ].
    \label{eq:appendix/probability exactly n}
\end{equation}

Using Eqs.\,\ref{eq:appendix/probability more than n} and \ref{eq:appendix/probability exactly n} it is possible to compute the probability that a galaxy has a given number of substructures or more. For example, one can check that Eq.,\ref{eq:appendix/probability of at least one clump} is recovered when plugging $n = 1$ into Eq.\,\ref{eq:appendix/probability more than n}. 

\end{document}